\documentclass[12pt,preprint]{aastex}
\def\degpoint{\ifmmode ^{\rm{o}}\!. \else $^{\rm{o}}\!.$\fi}

\newcommand{\ms}{\mbox{m\,s$^{-1}$}}
\newcommand{\kms}{\mbox{km\,s$^{-1}$}}
\newcommand{\Msun}{\mbox{M$_{\odot}$}}
\newcommand{\Rsun}{\mbox{R$_{\odot}$}}
\newcommand{\Mjup}{\mbox{M$_{\rm Jup}$}}

\newcommand{\ltsimeq}{\raisebox{-0.6ex}{$\,\stackrel
         {\raisebox{-.2ex}{$\textstyle <$}}{\sim}\,$}}
\newcommand{\gtsimeq}{\raisebox{-0.6ex}{$\,\stackrel
         {\raisebox{-.2ex}{$\textstyle >$}}{\sim}\,$}}

\begin{document}

\title{The Anglo-Australian Planet Search. XXII. Two New Multi-Planet 
Systems }

\author{Robert A.~Wittenmyer\altaffilmark{1}, J.~Horner\altaffilmark{1}, 
Mikko Tuomi\altaffilmark{2,3}, G.S.~Salter\altaffilmark{1}, 
C.G.~Tinney\altaffilmark{1}, R.P.~Butler\altaffilmark{4}, 
H.R.A.~Jones\altaffilmark{5}, S.J.~O'Toole\altaffilmark{6}, 
J.~Bailey\altaffilmark{1}, B.D.~Carter\altaffilmark{7}, 
J.S.~Jenkins\altaffilmark{8}, Z.~Zhang\altaffilmark{2}, 
S.S.~Vogt\altaffilmark{9}, Eugenio J.~Rivera\altaffilmark{9} }

\altaffiltext{1}{Department of Astrophysics, School of Physics, 
University of NSW, 2052, Australia}
\altaffiltext{2}{University of Hertfordshire, Centre for Astrophysics 
Research, Science and Technology Research Institute, College Lane, AL10 
9AB, Hatfield, UK}
\altaffiltext{3}{University of Turku, Tuorla Observatory, Department of 
Physics and Astronomy, V\"ais\"al\"antie 20, FI-21500, Piikki\"o, 
Finland}
\altaffiltext{4}{Department of Terrestrial Magnetism, Carnegie 
Institution of Washington, 5241 Broad Branch Road, NW, Washington, DC 
20015-1305, USA}
\altaffiltext{5}{Centre for Astrophysics Research, University of 
Hertfordshire, College Lane, Hatfield, Herts AL10 9AB, UK}
\altaffiltext{6}{Australian Astronomical Observatory, PO Box 296, 
Epping, 1710, Australia}
\altaffiltext{7}{Faculty of Sciences, University of Southern Queensland, 
Toowoomba, Queensland 4350, Australia}
\altaffiltext{8}{Departamento de Astronom\'ia, Universidad de Chile, 
Camino El Observatorio 1515, Las Condes, Santiago, Chile}
\altaffiltext{9}{UCO/Lick Observatory, University of California, Santa 
Cruz, CA 95064, USA}

\email{ 
rob@phys.unsw.edu.au}

\shorttitle{Two New Multi-Planet Systems }
\shortauthors{Wittenmyer et al.}

\begin{abstract}

\noindent We report the detection of two new planets from the 
Anglo-Australian Planet Search.  These planets orbit two stars each 
previously known to host one planet.  The new planet orbiting HD\,142 
has a period of 6005$\pm$427 days, and a minimum mass of 5.3\Mjup.  
HD\,142c is thus a new Jupiter analog: a gas-giant planet with a long 
period and low eccentricity ($e=0.21\pm0.07$).  The second planet in the 
HD\,159868 system has a period of 352.3$\pm$1.3 days, and 
m~sin~$i$=0.73$\pm$0.05 \Mjup.  In both of these systems, including the 
additional planets in the fitting process significantly reduced the 
eccentricity of the original planet.  These systems are thus examples of 
how multiple-planet systems can masquerade as moderately eccentric 
single-planet systems.

\end{abstract}


\section{Introduction}

Recent discoveries of multi-planet systems from maturing high-precision 
Doppler planet searches are revealing a surprising diversity of 
planetary system properties.  It is becoming apparent that, given a 
sufficient number of high-precision observations, many seemingly 
solitary stars or single-planet systems are found to host additional 
orbiting bodies.  This trend is evident both at extremely low masses 
(e.g.~Anglada-Escud\'e et al.~2012, Mayor et al.~2011, Vogt et al.~2010) 
and long periods (e.g.~HD~134987~c, Jones et al.~2010; 47~UMa~d, Gregory 
\& Fischer 2010).

Confirmation by independent observatories is extremely useful when 
testing the potential detection of additional planets, particularly if 
they have small radial-velocity amplitudes or the host star has a high 
level of velocity jitter.  For example, \citet{bean08} reported a third 
planet in the HD~74156 system using data from the Hobby-Eberly 
Telescope.  However, using the same spectra processed with two 
different, independent Doppler velocity codes, \citet{wittenmyer09} 
could not confirm that planet.  The Keck observations presented by 
\citet{mes11} were also inconsistent with a third planet in that system.  
High-precision and high-cadence data from multiple sites have proved 
critical in the confirmation of low-mass planets such as HD~4308 
\citep{udry06, otoole09}, the 61~Vir three-planet system \citep{61vir}, 
and HD~114613 \citep{tuomi12}.  For candidate multiple-planet systems, 
dynamical stability modelling is also a critical tool, since periodic 
signals arising from observational sampling or stellar activity can be 
misinterpreted as planets.  The inclusion of rigorous dynamical 
modelling has recently shown some candidate planetary systems to be 
unfeasible, e.g.~Horner et al.~(2011), Tuomi (2011), Wittenmyer et 
al.~(2012).

In this work, we present new data from the Anglo-Australian Planet 
Search (AAPS) which provide evidence for one additional planet orbiting 
HD~142 (\S 3.1) with a period of 6005 days.  We also perform a Bayesian 
analysis (\S 3.2), yielding results that are in agreement with the 
conventional least squares solution.  In Section 3.3, we present AAPS 
and Keck data indicating a second planet in the HD~159868 system with an 
orbital period of 352 days.  These two proposed multi-planet systems are 
subjected to detailed dynamical testing in Section~4.  In Section~5 we 
use direct imaging of the HD\,142 system to rule out the known stellar 
companion as the source of the observed radial-velocity signal.  Finally 
in Section~6, we place these discoveries in the context of the overall 
distribution of exoplanet properties and give our conclusions.

\section{Observations and Stellar Parameters}

AAPS Doppler measurements are made with the UCLES echelle spectrograph 
\citep{diego:90}.  Keck Doppler measurements are made with the HIRES 
spectrograph \citep{vogt94}.  An iodine absorption cell provides 
wavelength calibration from 5000 to 6200\,\AA.  The spectrograph 
point-spread function and wavelength calibration is derived from the 
iodine absorption lines embedded on every pixel of the spectrum by the 
cell \citep{val:95,BuMaWi96}.  The result is a precision Doppler 
velocity estimate for each epoch, along with an internal uncertainty 
estimate, which includes the effects of photon-counting uncertainties, 
residual errors in the spectrograph PSF model, and variation in the 
underlying spectrum between the iodine-free template, and epoch spectra 
observed through the iodine cell.  All velocities are measured relative 
to the zero-point defined by the template observation.

HD~142 has been observed by the AAT at 82 epochs, with a total data span 
of 5067 days.  HD~159868 has been observed by the AAT at 47 epochs, with 
a total data span of 3396 days.  In this analysis, we also add 34 Keck 
epochs spanning 1593 days.  The radial-velocity data for HD~142 are 
presented in Table~\ref{142vels}, and the data for HD~159868 are in 
Tables~\ref{159vels} and \ref{Keckvels}.  The physical parameters of 
HD~142 are given in \citet{tinney02}, and those for HD~159868 are given 
in \citet{otoole07}.  Tables~\ref{142star} and \ref{159star} summarise 
the parameters for HD~142 and HD~159868, respectively.  Briefly, HD~142 
is a G1IV star with a mass of 1.15$\pm$0.10 \Msun, is possibly slightly 
evolved \citep{tinney02}, is chromospherically inactive 
(log$R'_{HK}$=-4.95), with a moderately rapid rotation rate (V sin 
$i$=10.4 \kms).  HD~159868 is a G5 dwarf with a mass of 
1.087$^{+0.032}_{-0.033}$\Msun, solar metallicity, and is a 
chromospherically inactive (log$R'_{HK}$=-4.96) slow rotator (v sin 
$i$=2.1 \kms).  HD~142 has a known stellar companion \citep{poveda94}, 
which is a late K/early M star with a mass of $\sim$0.56\Msun\ 
\citep{egg07, rag06} and a projected separation of 105.1~AU.

\section{Orbit Fitting and Planetary Parameters}

\subsection{HD 142}

The Jupiter-mass planet orbiting HD~142 with a period of 339 days was 
one of the first discoveries reported by the AAPS \citep{tinney02}.  A 
further ten years of observations have revealed evidence for a 
long-period signal consistent with a second planet in the system.  In 
the discovery paper, the one-planet fit had a root-mean-square (RMS) 
residual scatter of 5.9 \ms, consistent with the 3--4 \ms\ estimated 
jitter for that star.  Using the formulation of \citet{wright05}, we now 
estimate a jitter of 4.5 \ms, which we apply in quadrature to the 
internal uncertainties shown in Table~\ref{142vels}.  A one-planet fit 
to the current data set for HD~142 now has an RMS of 31.1 \ms.  This 
significantly exceeds both the scatter due to the underlying precision 
of our AAPS Doppler measurement system and the predicted levels of 
stellar activity jitter expected for HD~142.  This excess scatter led us 
to investigate the possibility of one or more additional planets 
orbiting this star.

\begin{figure}
\plottwo{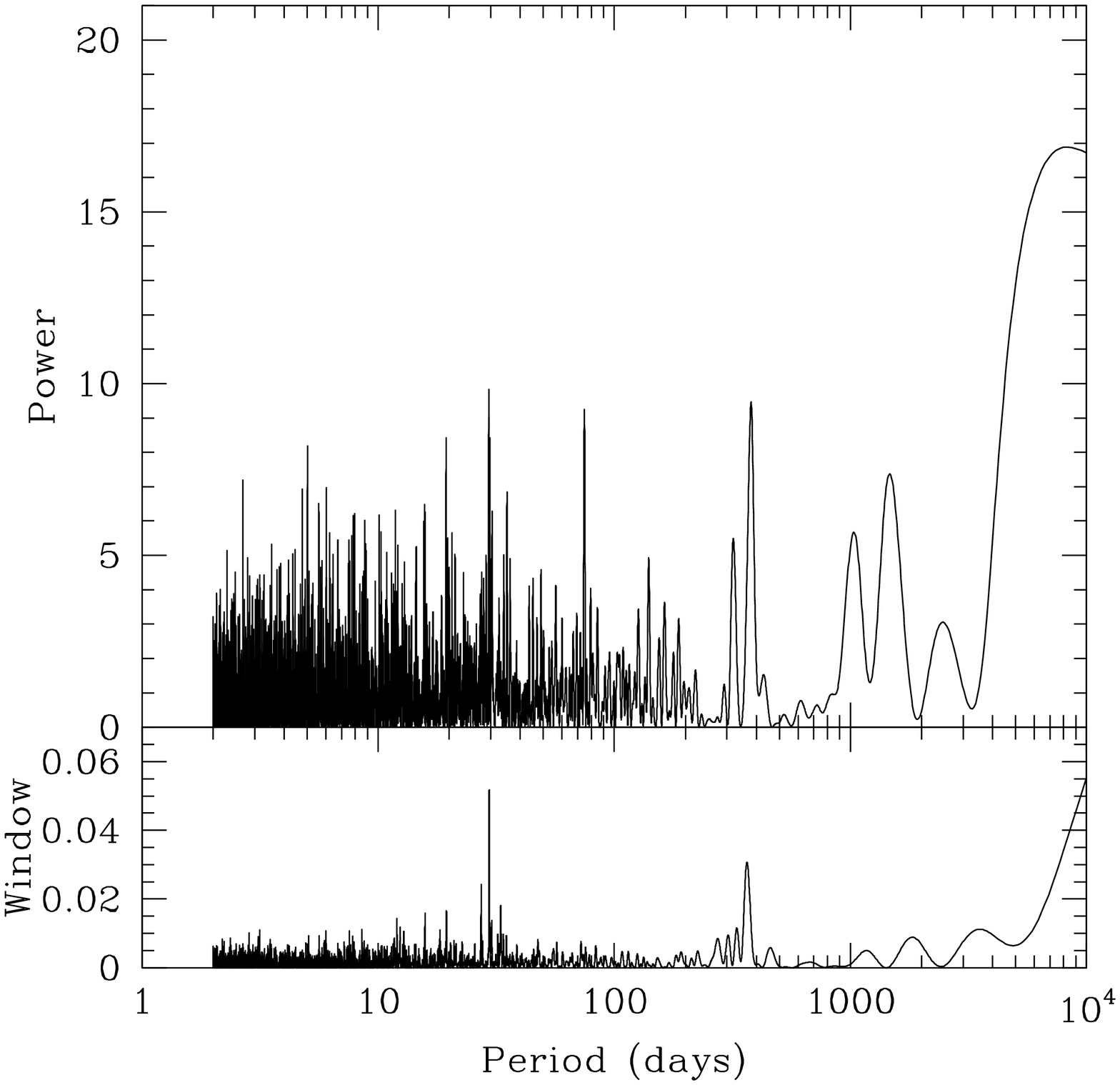}{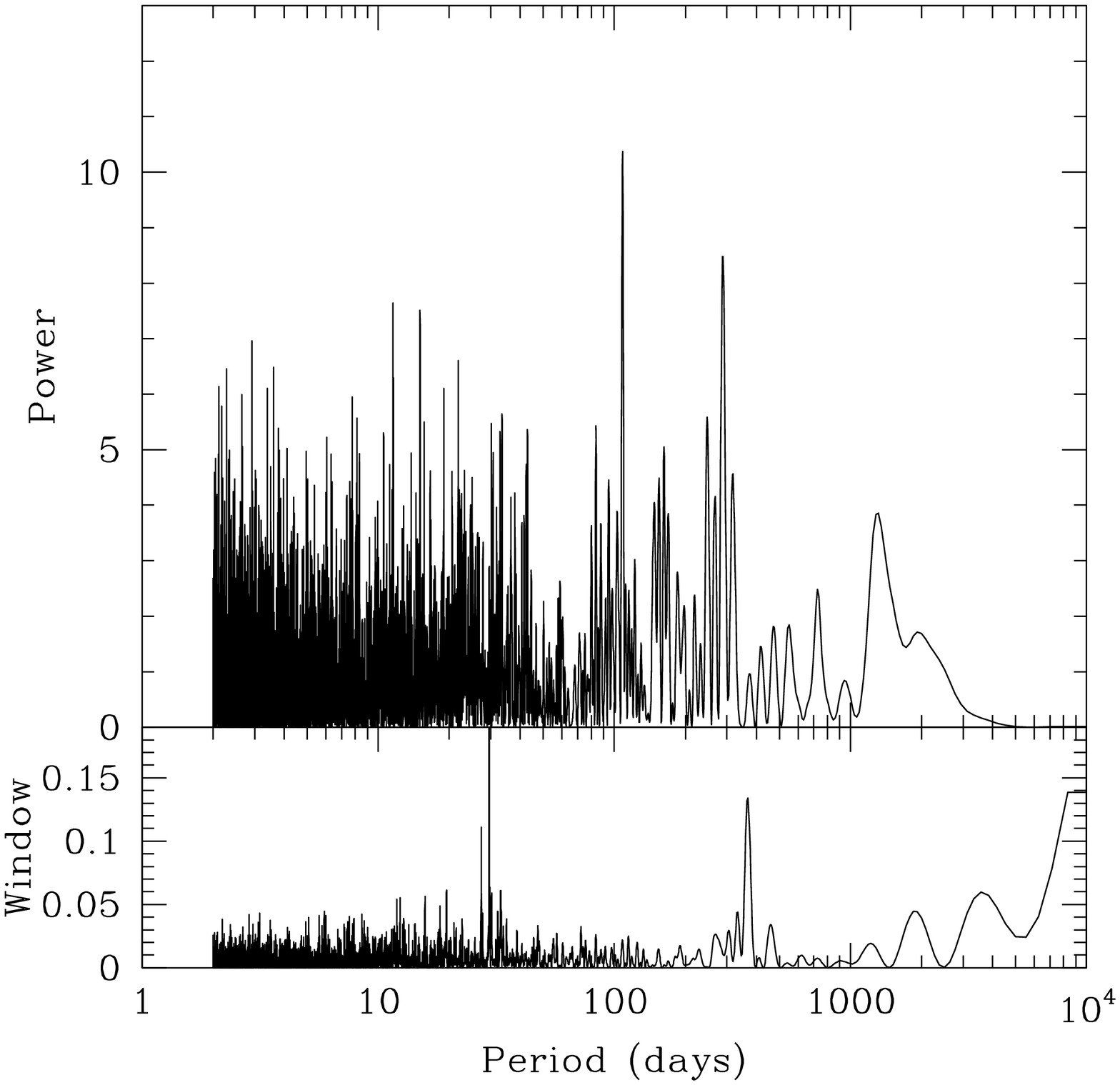}
\caption{Left panel: Periodogram of residuals for HD~142 after fitting 
one planet at $P=350$ days.  A long-period signal is clearly present.  
Right panel: Periodogram of residuals for HD~142 after fitting two 
planets, at $P=350$ and 6005 days.  The highest remaining peak is at a 
period of 108 days, with a false-alarm probability of 5.1\%. }
\label{142pgrams}
\end{figure}

Visual inspection of the residuals to the one-planet fit shows an 
obvious long-period signal ($P>1000$ days), and the periodogram shows a 
significant peak at very long periods (Figure~\ref{142pgrams}), so we 
proceed to fit a second planet.  First, we explored the vast and 
uncertain parameter space of the long-period signal with a genetic 
algorithm (e.g.~Cochran et al.~2007, Tinney et al.~2011, Wittenmyer et 
al.~2012).  We allowed the second planet to take on periods between 1000 
and 10000 days, and an eccentricity $e<0.6$.  The genetic algorithm ran 
for 50,000 iterations, each of which consisted of typically 1000--3000 
generations, during which the two-planet fits evolved toward a $\chi^2$ 
minimum.  The best-fit system parameters from this process are thus the 
result of $\sim 10^8$ trial Keplerian fits.  Used in this way, the 
genetic algorithm is an effective way of exploring a large parameter 
space, which is particularly important when the candidate planet's 
period is comparable to the length of the available data.  We then used 
the \textit{GaussFit} least-squares fitting code \citep{jefferys87} to 
obtain a Keplerian model fit, with the best 2-planet fit parameters from 
the genetic selection as initial inputs.

The two-planet fit has an RMS of 11.2 \ms, which is still somewhat 
higher than that expected for this star.  A periodogram of the residuals 
shows a peak at 108 days (Figure~\ref{142pgrams}).  We used a bootstrap 
randomisation process \citep{kurster97} to assess the false-alarm 
probability of the peak at 108 days.  The bootstrap method randomly 
shuffles the velocity observations while keeping the times of 
observation fixed.  The periodogram of this shuffled data set is then 
computed and its highest peak recorded.  From 10,000 such realisations, 
the peak at 108 days has a bootstrap false-alarm probability of 5.1\%.  
In mid-2011, this false-alarm probability was 2.5\% - that the addition 
of new data did not improve the statistical credibility of the 108-day 
signal leads us to conclude that the signal cannot be claimed as 
planetary in origin at this time.


The residual scatter about the two-planet fit remains much higher than 
expected given the jitter estimate of 4.45 \ms\ for HD~142.  However, 
the estimation of stellar activity jitter is a rather imprecise process, 
with uncertainties up to a factor of 2 (J.~Wright, priv.~comm.).  Hence, 
it is possible that we have underestimated the activity jitter for 
HD~142.  A jitter estimate of 11.3\,\ms\ is required to produce a 
reduced $\chi^2$ of unity for the two-planet fit.  Examining the 
distribution of jitter estimates for similar stars (Wright~2005, top 
panel of their Figure~7), the distribution has a tail extending toward a 
maximum jitter of 11\ms, but only includes 36 stars in total.  
\citet{if10} provide a different formulation to estimate activity 
jitter, using the Ca II S-index.  The Mount Wilson S-index for HD\,142 
is $S_{MW}=0.187$ \citep{jenkins06}.  Using the \citet{if10} 
formulation, this yields an estimated jitter of 3.07\ms.  Based on the 
chromospheric activity index and the high log\,$g$ 
(Table~\ref{142star}), it appears that HD\,142 is an inactive dwarf 
star, which nonetheless presents a high level of radial-velocity noise.  
We note that this star is quite a rapid rotator: it has a v~sin\,$i$ of 
10.4 \kms, compared to typical planet-search targets which have 
v~sin\,$i$ of 2-4 \kms.  We conclude that the poor velocity precision 
for HD\,142 is attributable to the rapid rotation, which broadens the 
spectral lines and limits our ability to derive extremely precise radial 
velocities.

\begin{figure}
\plotone{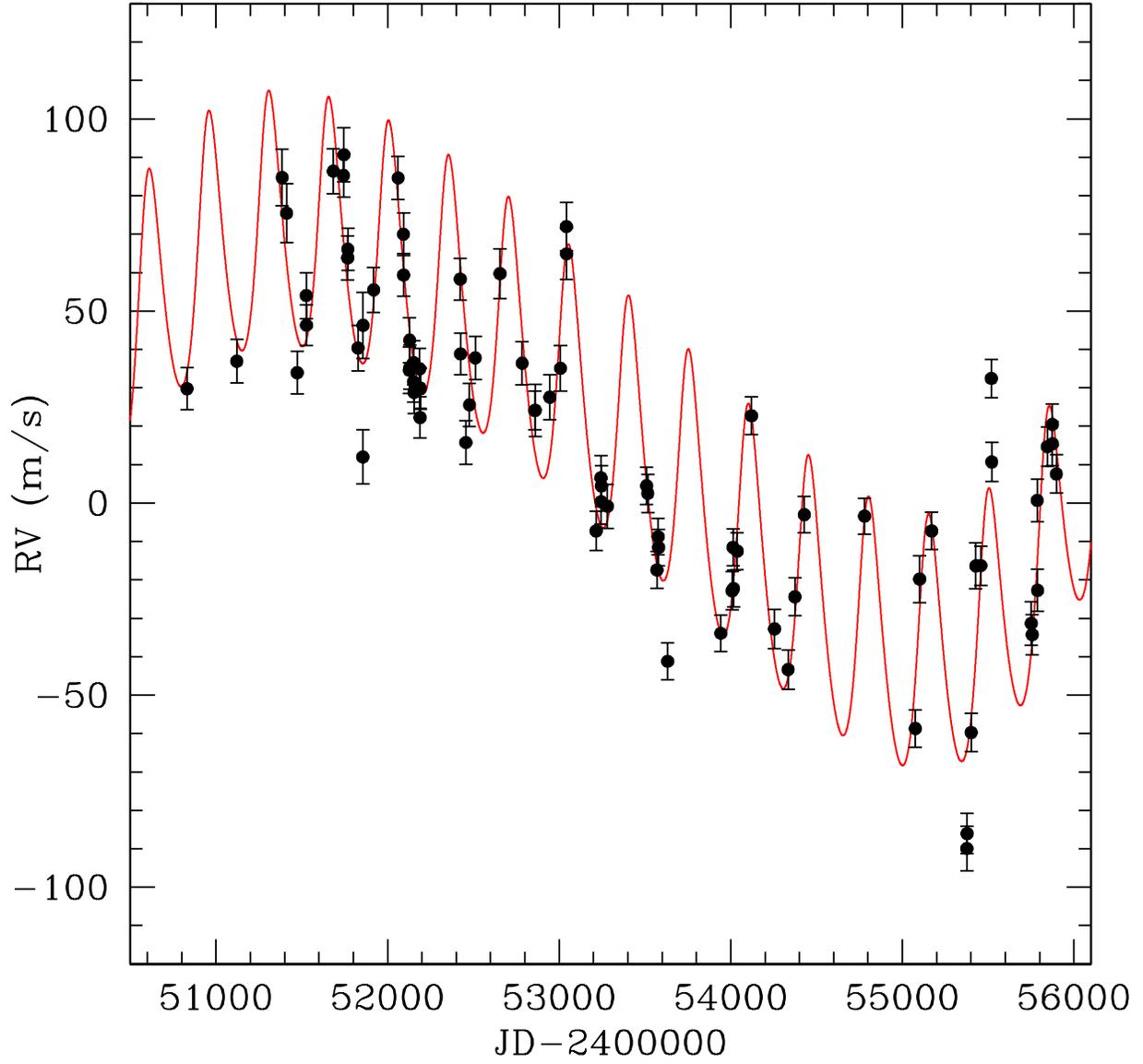}
\caption{Two-planet fit for HD~142.  The residuals of this fit are 11.2 
\ms, and no further significant signals are present. }
\label{142fit}
\end{figure}
\begin{figure} 
\plottwo{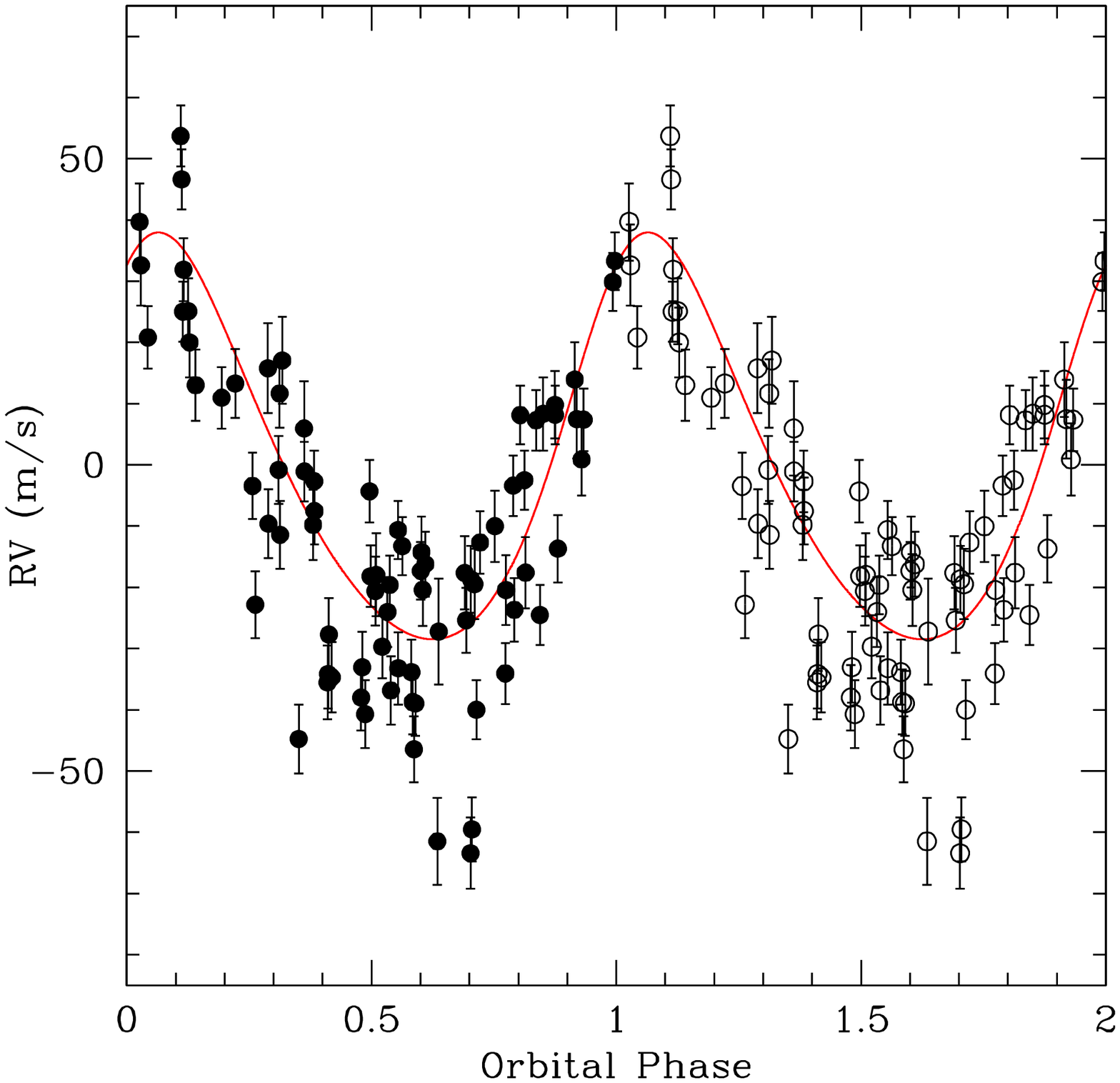}{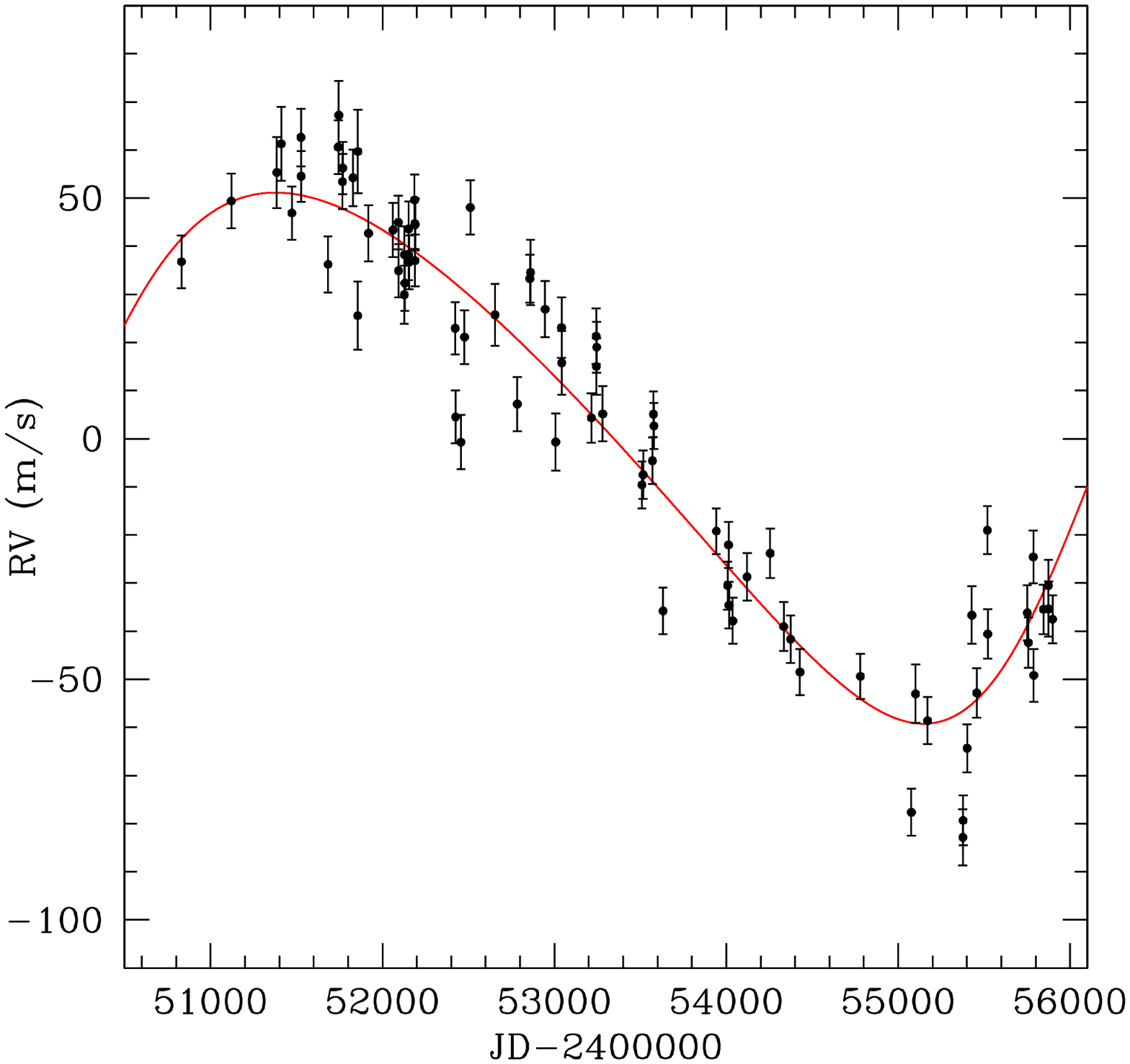} 
\caption{Left panel: Fit for HD~142b, the previously known planet with 
$P=350$ days. Two cycles are shown for clarity.  Right panel: Fit for 
HD~142c, with $P=6005$ days.  In both panels, the signal of the other 
planet has been removed.}
\label{more142fits} 
\end{figure}

Using a stellar mass of 1.15$\pm$0.10 \Msun\ \citep{tinney02}, we 
estimate the minimum mass, m~sin~$i$, for the outermost planet 
(planet~c) to be 5.3$\pm$0.7 \Mjup.  The 2-planet fit is shown in 
Figure~\ref{142fit} and the planetary parameters are given in 
Table~\ref{planetparams}.  The individual fits for each of the two 
planets are shown in Figure~\ref{more142fits}.

\subsection{A Bayesian Analysis for HD 142}

Thirteen years of AAT data have provided evidence for a very long-period 
planet (HD~142c) with a period of more than 6000 days.  The baseline of 
these AAT observations is the longest currently available at high 
precision for this star, which makes independent confirmation of HD~142c 
problematic.  A third candidate signal is present, with a period of 108 
days and a velocity semiamplitude $K\sim12$\ms, which is comparable to 
the 11.2\ms\ residual velocity scatter about the 2-planet fit.  It is 
prudent, then, to employ an independent analysis to test the 
plausibility of the 108-day signal.

We analysed the AAPS radial velocities using posterior samplings of 
different models and the comparisons of these models using Bayesian 
model probabilities \citep[e.g.][]{tuomi2009,tuomi2011,tuomi2011b}.  We 
used the adaptive Metropolis algorithm \citep{haario2001} for posterior 
samplings because it appears to be a reasonably efficient method for 
analysing radial velocities with Keplerian models 
\citep{tuomi2011,tuomi2011b}.  We present the results using the maximum 
\emph{a posteriori} (MAP) estimates and the corresponding 99\% 
credibility intervals ($\mathcal{D}_{0.99}$), i.e. the Bayesian 
credibility sets as defined in \citet{tuomi2009}.  Our prior probability 
densities of the model parameters are those used in \citet{ford2007}, 
with slight modifications. We penalised very high eccentricities by 
setting the prior densities for orbital eccentricities $\pi(e) \propto 
\mathcal{N}(0, \sigma_{e}^{2})$, where the parameter $\sigma_{e}$ was 
set to have a value of 0.3 that still allows the orbital eccentricities 
to have high values if the data insists so.  We also adopted 
conservative prior probabilities for models with $k$ Keplerian signals 
such that $P(\mathcal{M}_{k}) = 2 P(\mathcal{M}_{k+1})$, i.e.~that the 
prior probability of having $k+1$ planets in the system is always two 
times less than having $k$ planets. Essentially, this enables us to be 
more confident with the interpretation of our model probabilities -- if 
there appear to be $k$ Keplerian signals in the data, we actually 
underestimate the significance of the weakest signal because of these 
priors.

The probabilities of models $\mathcal{M}_{k}, k=0, ..., 3$ planets and 
the corresponding RMS values are shown in 
Table~\ref{model_probabilities}.  The velocity jitter was also included 
as a free parameter in the Bayesian model, and the best-fit jitter 
values are given in Table~\ref{model_probabilities} for each of the 
$k$-planet models.  These results support the presence of three 
Keplerian signals in the data, although the parameters of the 108-day 
signal (``planet d'') are poorly constrained.  The orbital parameters of 
this three-planet solution are shown in Table~\ref{HD142_parameters}.  
The distribution of allowed orbital periods for the outer planet had a 
substantially longer tail toward longer periods (as expected given the 
data span is shorter than the expected orbital period for this planet).  
The large uncertainty in the period of planet c maps directly into large 
uncertainties for the eccentricity and velocity semiamplitude $K$.  We 
therefore note that the nominal 1$\sigma$ uncertainties in the 
parameters of planet c as given in Table~\ref{planetparams} are likely 
underestimated, and we advise the reader to consider the 99\% confidence 
intervals in Table~\ref{HD142_parameters} as more comprehensive.


\subsection{HD 159868}

\citet{otoole07} reported the detection of a long-period ($P=986$ days), 
eccentric ($e=0.69$) planet orbiting HD~159868, based on 4.5 years of 
AAT data.  That fit had a residual RMS scatter of 8.4 \ms, which the 
authors noted as larger than expected for that star.  They speculated 
that a second planet with $P=180$ days would substantially reduce the 
RMS but, wary of the sampling difficulties in constraining planet 
candidates with periods near one-half of a sidereal year, they presented 
only the single-planet solution for HD~159868.  Now, with nearly twice 
as much data (Table~\ref{159vels}), the single-planet fit has an RMS of 
15.8 \ms.  As is the case for HD\,142 above, the worsening single-planet 
fit gives a clue that additional planets are present in this system.  
Here we adopt a jitter estimate of 2.65 \ms, and apply this in 
quadrature to the uncertainties given in Table~\ref{159vels} before 
performing orbital fitting.  The fitting procedures used followed those 
outlined above for HD\,142.

\begin{figure}
\plotone{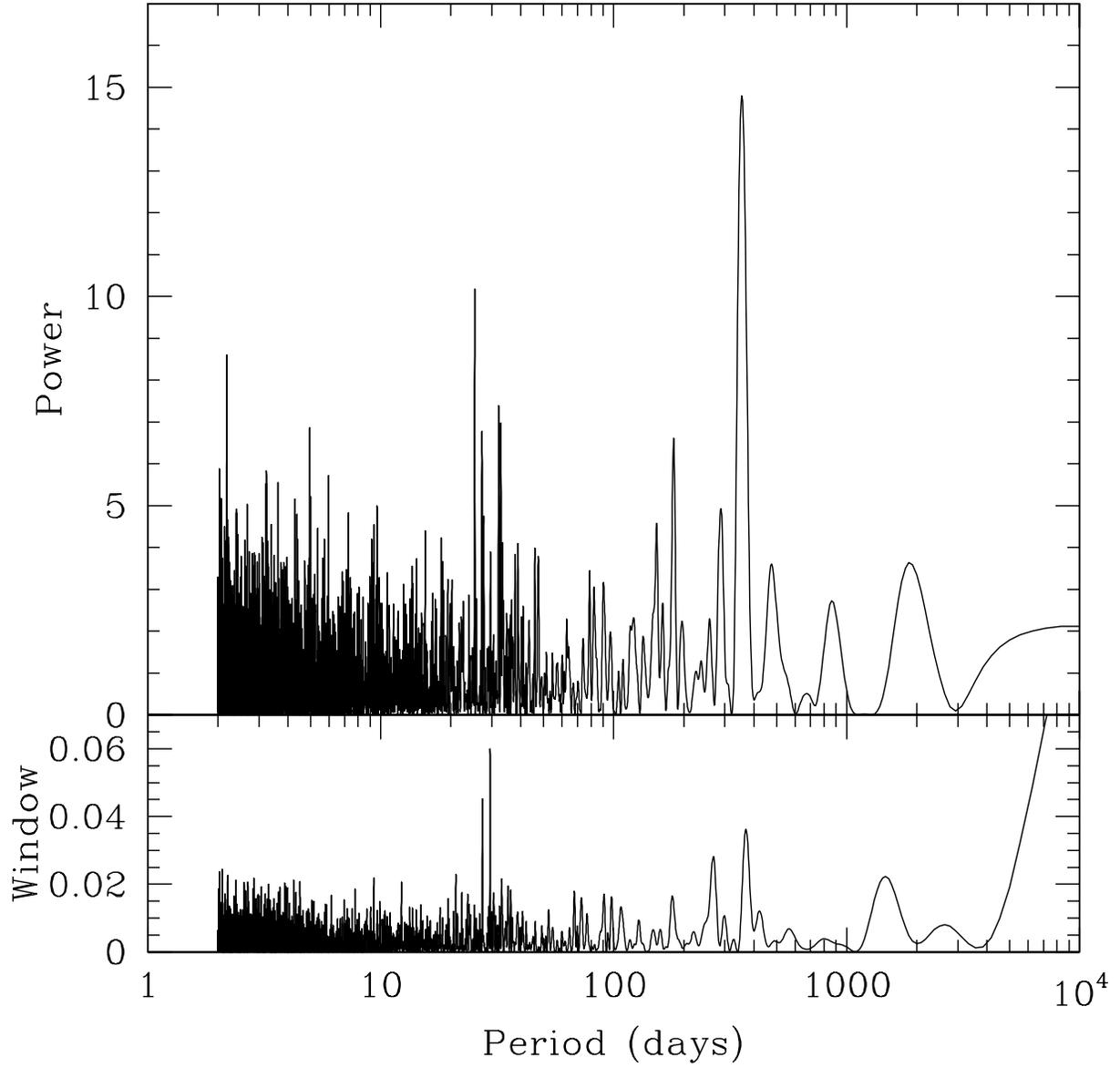}
\caption{Periodogram of AAT residuals for HD~159868 after fitting one planet 
at $P=1178$ days.  An additional signal is present near 355 days. }
\label{159pgram}
\end{figure}

\begin{figure}
\plottwo{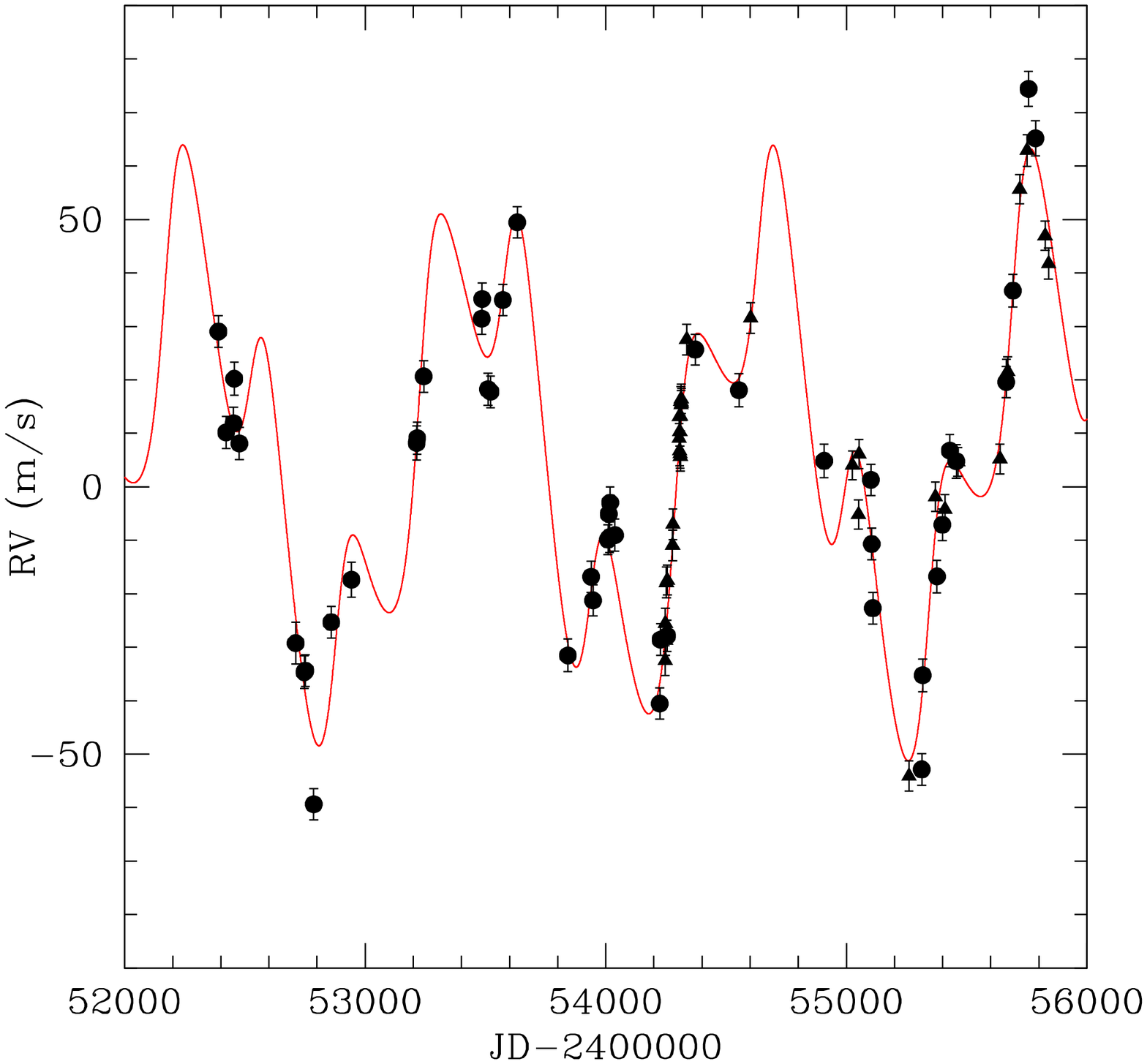}{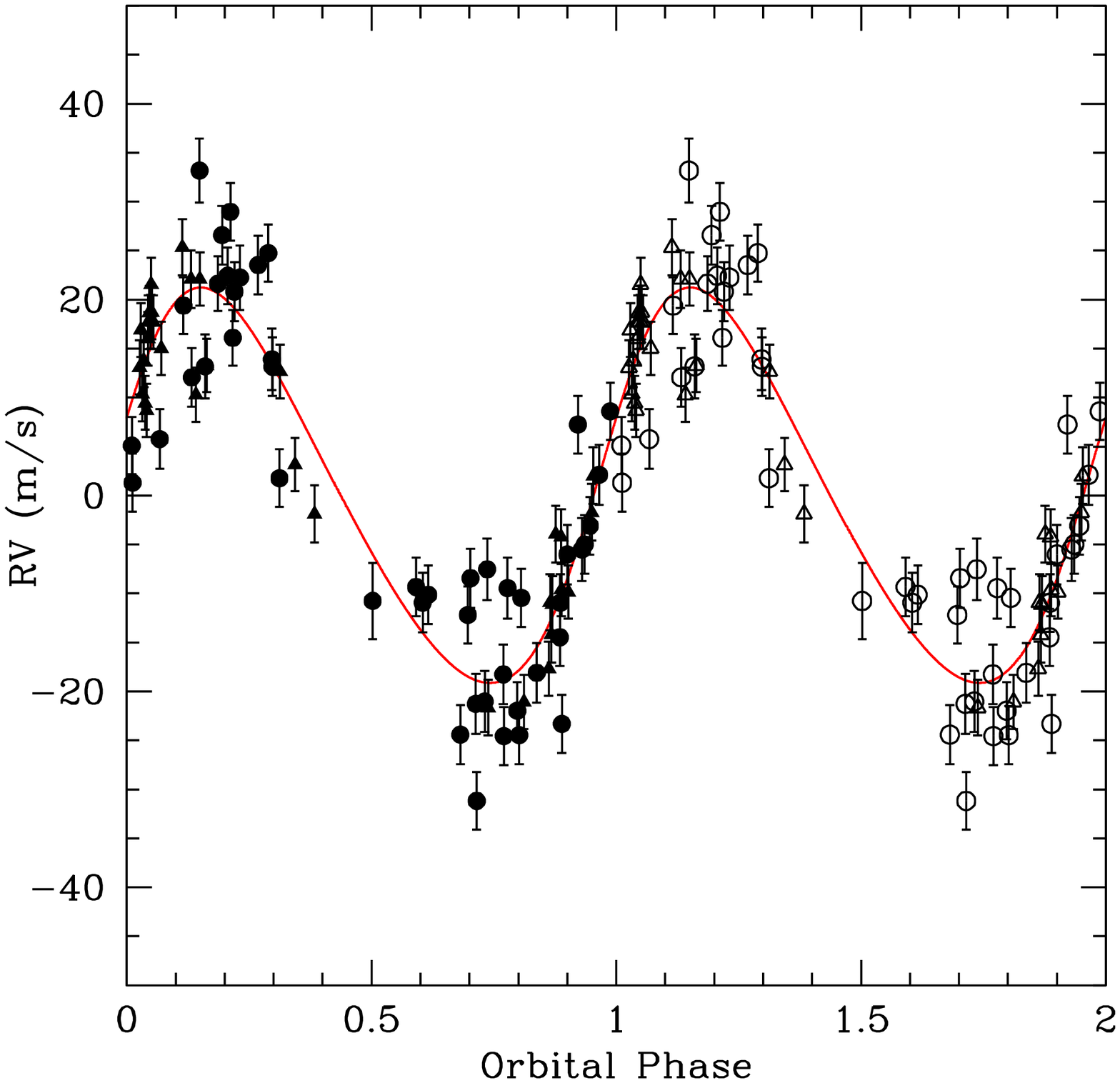}
\caption{Left panel: Two-planet Keplerian fit for HD~159868; Circles are 
AAT data, and triangles are Keck data.  The total RMS about of this fit 
is 5.8 \ms.  Right panel: Fit for the 355-day planet only, folded to 
show phase coverage.  The symbols have the same meaning, and two cycles 
are shown for clarity. }
\label{159fit}
\end{figure}

Interestingly, a 1-planet fit for HD~159868 now has an eccentricity of 
only 0.16$\pm0.11$, which is \textit{markedly} different from the 
$e=0.69$ solution presented in \citet{otoole07}.  A periodogram of the 
residuals to this fit (Figure~\ref{159pgram}) shows a large peak at 355 
days.  This peak has a bootstrap false-alarm probability $<$0.01\%.  Now 
armed with substantial evidence for a second planet in the system, we 
proceed with a 2-Keplerian solution.  For the final orbit fit, we also 
include 4.3 years of data from the Keck telescope.  The best fit has a 
second planet with a period of 352.3$\pm$1.3 days 
(Table~\ref{planetparams}); adopting a stellar mass of 
1.087$^{+0.032}_{-0.033}$ \citep{takeda07}, the planet has 
m~sin~$i=0.73\pm$0.05 \Mjup.  Figure~\ref{159fit} shows the 2-planet fit 
and the phase coverage for the new planet candidate, which has a period 
near one year.  This fit has a total residual RMS of 5.8 \ms\ (AAT -- 
6.7 \ms; Keck -- 4.6 \ms), which is still somewhat higher than expected 
based on the instrumental noise and stellar activity jitter.  Based on 
the somewhat low log\,$g$ values given in Table~\ref{159star}, 
HD\,159868 may be slightly evolved.  If HD\,159868 is a subgiant, the 
velocity jitter may be closer to $\sim$5\ms, typical of subgiants 
\citep{kb95, johnson10a, 47205paper} and consistent with the RMS scatter 
about our two-planet fit.

A periodogram of the residuals now has a peak at 12.6 days, but it is 
not significant, with a bootstrap false-alarm probability of 27\%.  As a 
further check, we also performed a Bayesian analysis as described above 
for the HD\,142 system, and we obtain model probabilities 
(Table~\ref{bayes159}) which confidently indicate two Keplerian signals.  
These results demonstrate how two planets in nearly-circular orbits can 
mimic a single eccentric planet when data are sparse and more subject to 
vagaries of sampling.  This can cause signals near one year to be missed 
\citep{tinney11, ang10}.

\section{Dynamical Stability Analysis}


While single planet systems can be fully solved with a simple Keplerian 
analysis, gravitationally interacting systems of multiple planets 
require a full Newtonian analysis.  Proposed solutions need to be shown 
to be dynamically stable over reasonably long timescales 
\citep{horner11, HUAqr, HUAqr2} to be considered ``real.'' 
Gravitationally interacting systems provide both an independent check on 
``reality'' \citep{fab12}, and a useful means to constrain or solve for 
the orbital inclination angle and true mass of the planets 
\citep{rivera05}.

\subsection{The HD 142 System}

To examine the dynamical stability of the proposed HD\,142 planetary 
system, we performed a series of highly detailed $n$-body dynamical 
simulations of a wide range of potential system architectures. Given the 
extreme uncertainty in the orbit of the outermost planet detected in the 
HD\,142 system, we concentrated solely on the orbital stability of the 
350-day planet and a possible 108-day planet.  As discussed in Section 
3.1, there is a residual signal after fitting two planets; the 
false-alarm probability of that signal is not presently low enough to 
justify claiming a third planet.  However, if further observational data 
support the existence of such an object, it would be wise to understand 
the dynamics of the system.  In this subsection, we explore the 
dynamical interactions between the known 350-day planet and the possible 
108-day planet.


Following previous studies of exoplanetary stability \citep{marshall10, 
horner11, HUAqr, HUAqr2, Texan}, we used the Hybrid integrator within 
the $n$-body dynamics package \textit{MERCURY} \citep{chambers99} to 
examine the stability of the planetary system as a function of its 
orbital architecture.  Following those earlier works, we considered 
two-planet systems in which planet~b (the most well-constrained) was 
placed on its nominal best-fit orbit (Table~\ref{planetparams}).  The 
best fit for a potential third planet has $P=108.2\pm$0.2 days, 
$K=12.3\pm$1.9 \ms, $e=0.28\pm$0.16, and $\omega = 271\pm$27 degrees.  
The initial orbital elements were then uniformly distributed across the 
3$\sigma$ confidence range in $a$, $e$, and mean anomaly $M$.  In total, 
we tested 35 values of $a$, 35 values of $e$, and 25 values of $M$, 
creating a grid of 30,625 initial system architectures.  The semi-major 
axes for the innermost planet varied between 0.44 and 0.50, the 
eccentricity was varied between 0.00 and 0.76, and the mean anomalies 
were varied between 5 and 355 degrees.  Each of these systems was then 
integrated for a period of 100 million years.  A body was considered to 
be ejected from the system if it reached a distance of 10~AU from the 
central star.  For each of the 30,625 potential systems, we obtained 
either the time at which the system fell apart (through collisions or 
ejections), or alternatively found that the system remained intact until 
the end of our simulations.

\begin{figure}
\plotone{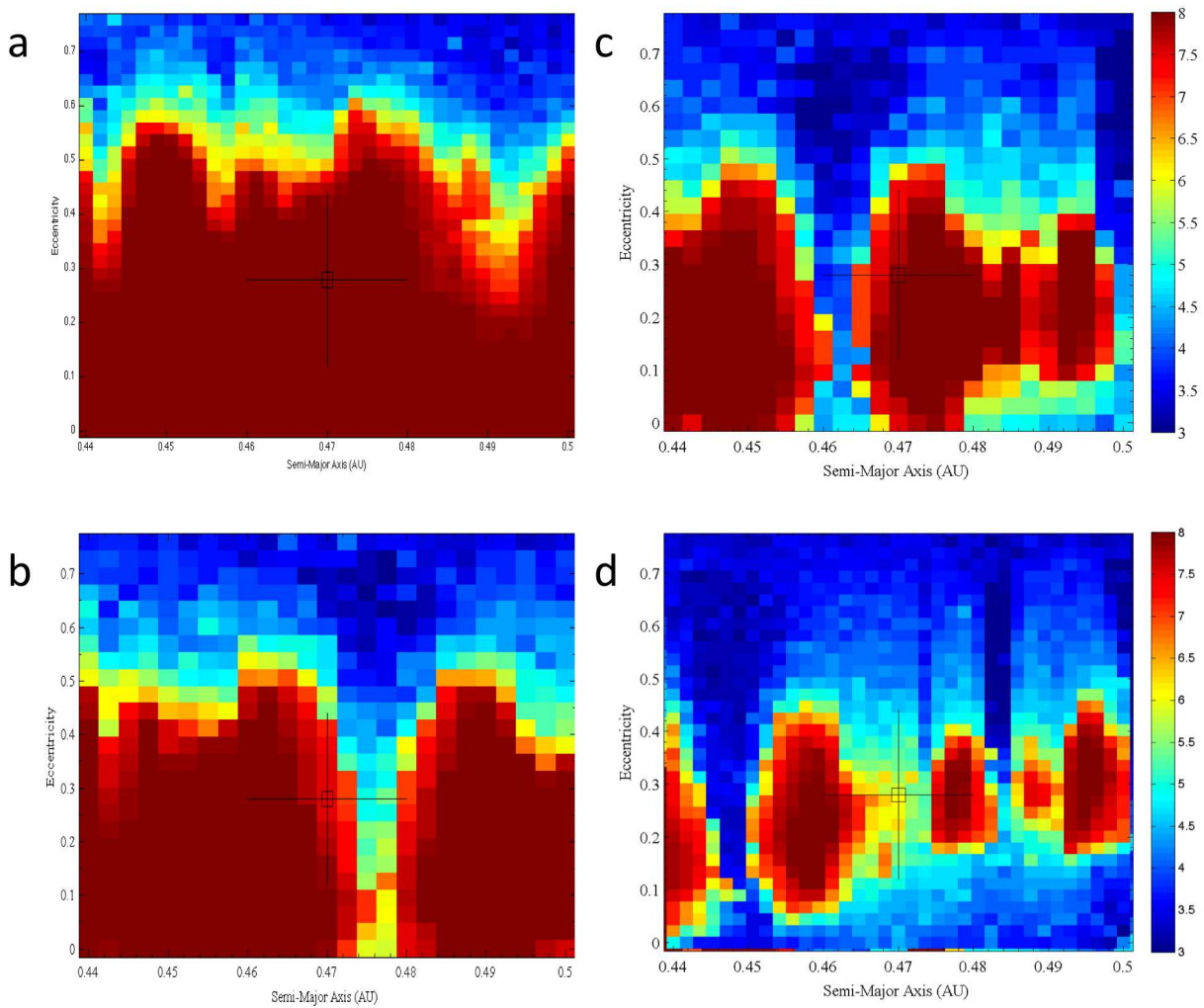}  
\caption{Dynamical stability of the HD~142 system, where planet~b is 
fixed on its best-fit orbit (panel a), and a candidate planet d is 
present at $a=0.47$AU.  The color bar indicates the survival time in log 
years.  Subsequent panels give the results when planet~b is fixed on a 
higher eccentricity and smaller semimajor axis, in steps of $1\sigma$.  
Each square in the grid shows the mean lifetime of 25 independent 
simulations (9, for panels b and c).  Panels (b)--(d) show the 
increasing instability as various mean-motion resonances (particularly 
the 3:1) move with planet~b.}
\label{142Dynam4plot}
\end{figure}

These results allow us to construct detailed dynamical maps of the 
system.  Figure~\ref{142Dynam4plot} shows the results: detailed 
dynamical maps of the HD\,142 system.  Each panel shows the mean 
lifetime of the system, as a function of semi-major axis and 
eccentricity, with each coloured square showing the mean of the 
twenty-five individual runs carried out at that particular $a$-$e$ 
location.  Panel (a) shows the results using the nominal best-fit orbit 
for planet~b: the entire region spanned by the $\pm\,1\sigma$ 
uncertainties on the orbit of planet~d is dynamically stable.  The only 
departures from stable solutions are found at relatively large orbital 
eccentricities for planet~d, which cause the two planets to experience 
mutually destabilising encounters.

Following these test integrations, we examined the most extreme case 
possible within the 3$\sigma$ error bounds on the orbit of planet~b.  We 
repeated the integrations exactly as described above, but placed 
planet~b on the most eccentric orbit allowed within the 3$\sigma$ 
confidence interval, and at the smallest semi-major axis that interval 
would allow (i.e.~$a=0.93$ AU, $e=0.37$).  This was designed to give the 
system the greatest possible chance of instability -- essentially to 
test it to destruction.  These results are shown in panel (d) of 
Figure~\ref{142Dynam4plot}.  In contrast to the integrations described 
above, the great majority of the phase space tested for this extreme 
scenario turns out to be unstable, although there remain several broad 
``islands of stability.''  These islands are separated by a wealth of 
unstable regions, primarily driven by the web of mutual mean-motion 
resonances between the two planets.

To better illustrate how the region of instability varies as a function 
of the orbit of a potential 108-day planet ``d,'' we carried out two 
subsidiary suites of integrations.  These again covered the full 
$\pm\,3\sigma$ range of $a$-$e$ space for planet~d, but with a 
resolution of 25x25x9 steps in $a$-$e$-$M$.  First, we placed planet~b 
on an orbit with eccentricity $1\sigma$ greater than the nominal value, 
and a semi-major axis $1\sigma$ smaller than the nominal value (i.e. 
$a=0.99 AU$, $e=0.27$).  In the second suite, planet~b was placed on an 
orbit $2\sigma$ more eccentric than, and $2\sigma$ inside, the nominal 
values (i.e. $a=0.96 AU$, $e=0.32$).  These results are shown in panels 
(b) and (c) of Figure~\ref{142Dynam4plot}, respectively.  Thus, panels 
(b)--(d) show the results when the orbital eccentricity and semi-major 
axis of planet~b are changed in $1\sigma$ steps from their nominal 
values.

It is clear that the scenarios featured in panels (a) and (b) of 
Figure~\ref{142Dynam4plot} reveal a far greater proportion of 
dynamically stable orbits.  One obvious result from these dynamical 
tests is that when the eccentricity of a candidate planet d increases, 
the stability of the system dramatically decreases.  The cross at the 
center of each panel in Figure~\ref{142Dynam4plot} shows the best fit 
and 1$\sigma$ uncertainties for a third planet.  The most recent data 
for HD\,142 now result in a higher eccentricity for a third planet 
($e=0.4\pm$0.1).  The results of our dynamical simulations, in which 
higher eccentricities for a proposed planet~d were less likely to remain 
stable, are in agreement with the analysis in Section 3.1: a 108-day 
planet in this system is increasingly unlikely.


\subsection{The HD 159868 System}

To test the orbital stability of the two planets discovered in the HD 
159868 system, we once again performed two detailed suites of dynamical 
integrations of the planetary system, using the Hybrid integrator within 
\textit{MERCURY}.  In the first suite, as for the first set of runs 
performed for the planets in the HD\,142 system, we considered scenarios 
in which HD159868~b was placed on its nominal orbit ($a = 2.25$ AU, $e = 
0.05$, etc.).  We then carried out 30,625 individual simulations, 
through which the orbit of HD\,159868~c was varied across the full 
3-$\sigma$ range of allowed orbital solutions in $a$, $e$ and $M$ (in a 
35x35x25 grid, again as before). In stark contrast to the results for 
HD\,142, which featured a significant number of unstable solutions, 
every single system tested for HD\,159868 remained dynamically stable 
for the full 100 Myr of our study.

Following that first suite of integrations, we carried out a second 
test, in which the orbit of HD159868~b was set to the most extreme 
allowed within the 3-$\sigma$ uncertainties, with $a = 2.13$ AU and $e = 
0.17$. Once again, we tested 30,625 unique planetary systems, with the 
orbit of HD\,159868~c varied across the full 3-$\sigma$ range of allowed 
orbital solutions.  These extreme runs did yield a small number of 
unstable solutions (162 of the 30,625 runs were destabilised by the end 
of the integrations), but the vast majority of systems tested survived 
unscathed until the end of the simulations. Every unstable solution 
required the initial orbit of HD\,159868~c to have an eccentricity of at 
least 0.32, and almost all featured initial semi-major axes of greater 
than 1.032 AU.

\section{Direct Imaging for HD 142}

Since the outermost object in the HD\,142 system has a very long and 
poorly-constrained orbital period, it is prudent to check whether the 
observed radial-velocity variation is due to a stellar companion on a 
much longer-period orbit.  As noted in Section~2, HD\,142 is known to 
host a stellar companion (0.56\Msun) with a projected separation of 
105.1~AU.  If this companion were the source of the radial-velocity 
signature attributed to the outermost companion, it would require the 
system to be almost face on (an inclination angle of $\sim0.11$ degrees 
for the derived $e=0.20$ and $K=55.5$\ms).

\begin{figure}
\plotone{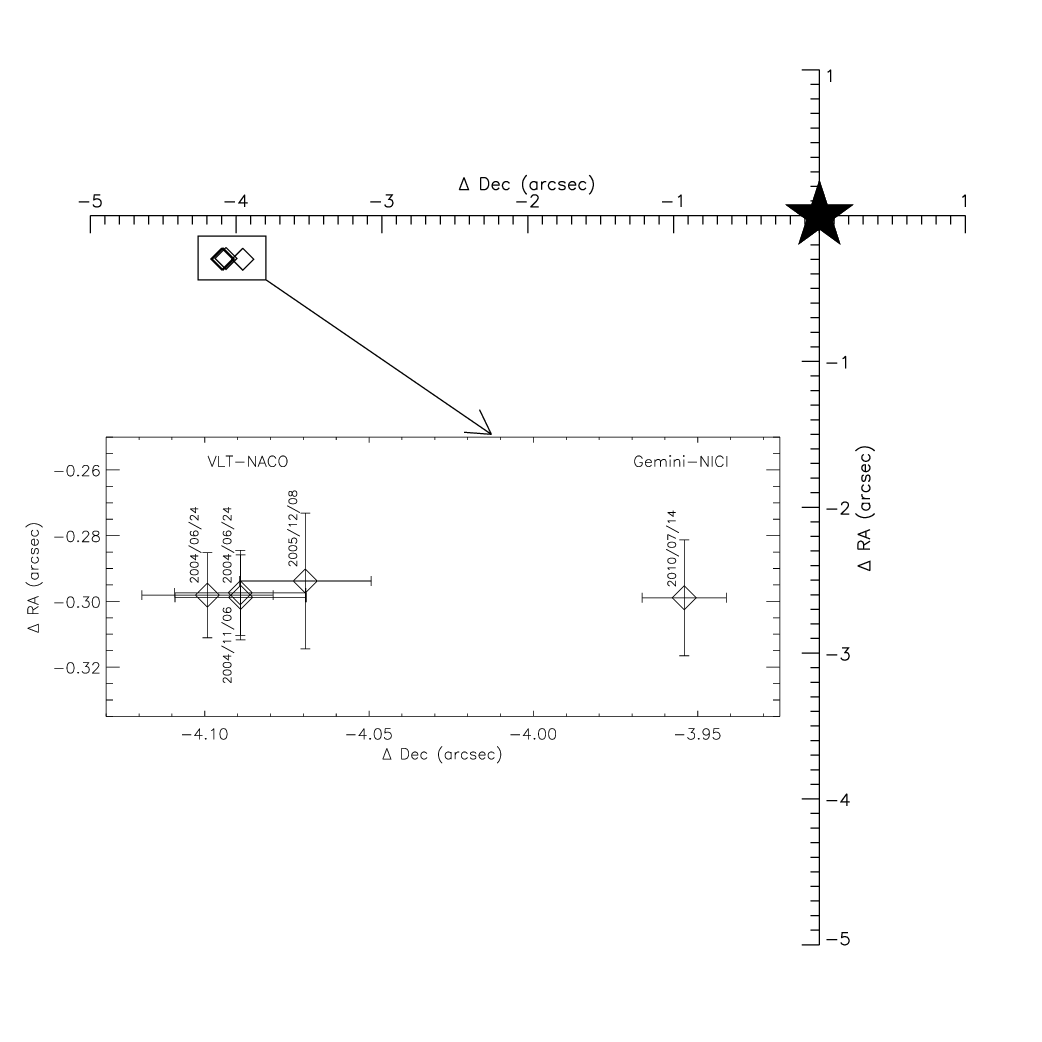}
\caption{Observed separations between HD\,142 and its stellar companion. 
The 4 data points on the left hand side are from VLT-NACO \citep{egg07}, 
and the solitary data point on the right hand side is from Gemini-NICI 
observations. The date of observation is indicated next to each data 
point. Further details can be found in Table~\ref{nici}. }
\label{imaging}
\end{figure}

Combining previous VLT-NACO observations of the HD\,142 system, 
previously published in \citet{egg07}, and a more recent observation we 
made with the Near-Infrared Coronographic Imager (NICI) on the 8m Gemini 
Telescope (Figure~\ref{imaging} and Table~\ref{nici}), we can clearly 
see the stellar companion moving almost directly towards the star which 
suggests a nearly edge-on system.  The stellar companion seen from 
direct imaging would also have an orbital period of $>1000$ yr instead 
of the 6005 days (17 yr) determined from the radial-velocity data.  This 
evidence leads us to believe that the stellar companion is not likely to 
be the cause of the radial-velocity variation attributed to HD\,142c.

\section{Discussion}

In \citet{jupiters}, we defined a Jupiter analog as a gas-giant planet 
with a period $P\gtsimeq$8 yr and a small eccentricity ($e\ltsimeq$0.2).  
HD\,142c has a period of 17 years, $e=0.2$, and a mass estimate 
consistent with a gas-giant planet.  This planet thus represents a new 
Jupiter analog, the fourth such planet discovered by the AAPS.  The 
three previous AAPS Jupiter analogs are HD\,134987c \citep{jones10}, 
GJ\,832b \citep{bailey09}, and HD\,160691c \citep{mccarthy04}.  This 
discovery of a new, very long-period planet provides additional evidence 
that continued support for the AAPS is bearing fruit.  We note that 
although the best-fit period for HD\,142c is longer than the duration of 
observations, the available data cover $\sim$80\% of an orbital cycle.  
\citet{jupiters} investigated the extant literature and found that the 
minimum orbital coverage for published planets was 70\% of an orbital 
cycle.

The AAPS data also show hints of a third signal with a period of 108 
days, evident in both the traditional periodogram analysis and the 
Bayesian analysis (\S 3.2).  Since this signal still has a false-alarm 
probability of 5.1\%, we do not claim it to be a planet at this time.  
We note that while very low-amplitude planets have been detected, with 
amplitudes comparable to the stellar radial-velocity noise (e.g.~Vogt et 
al.~2010, Pepe et al.~2011), those detections were clearly evident in 
the periodograms, and the host stars were slow rotators with extremely 
low intrinsic jitter (unlike HD\,142).  


These discoveries also highlight the importance of continuing to monitor 
known planetary systems for signs of additional objects.  In particular, 
both HD\,142 and HD\,159868 had moderately eccentric orbital solutions 
as well as excess scatter about the single-planet fit.  A dedicated 
search pursuing this strategy was performed by \citet{wittenmyer09}, who 
observed 22 known planetary systems for 3 years using the Hobby-Eberly 
Telescope.  While that survey did not result in new planet dscoveries, 
the new data cast doubts on the existence of the proposed planets 
HD~20367b \citep{udry03} and HD~74156d \citep{bean08, mes11}.

\begin{figure}
\plotone{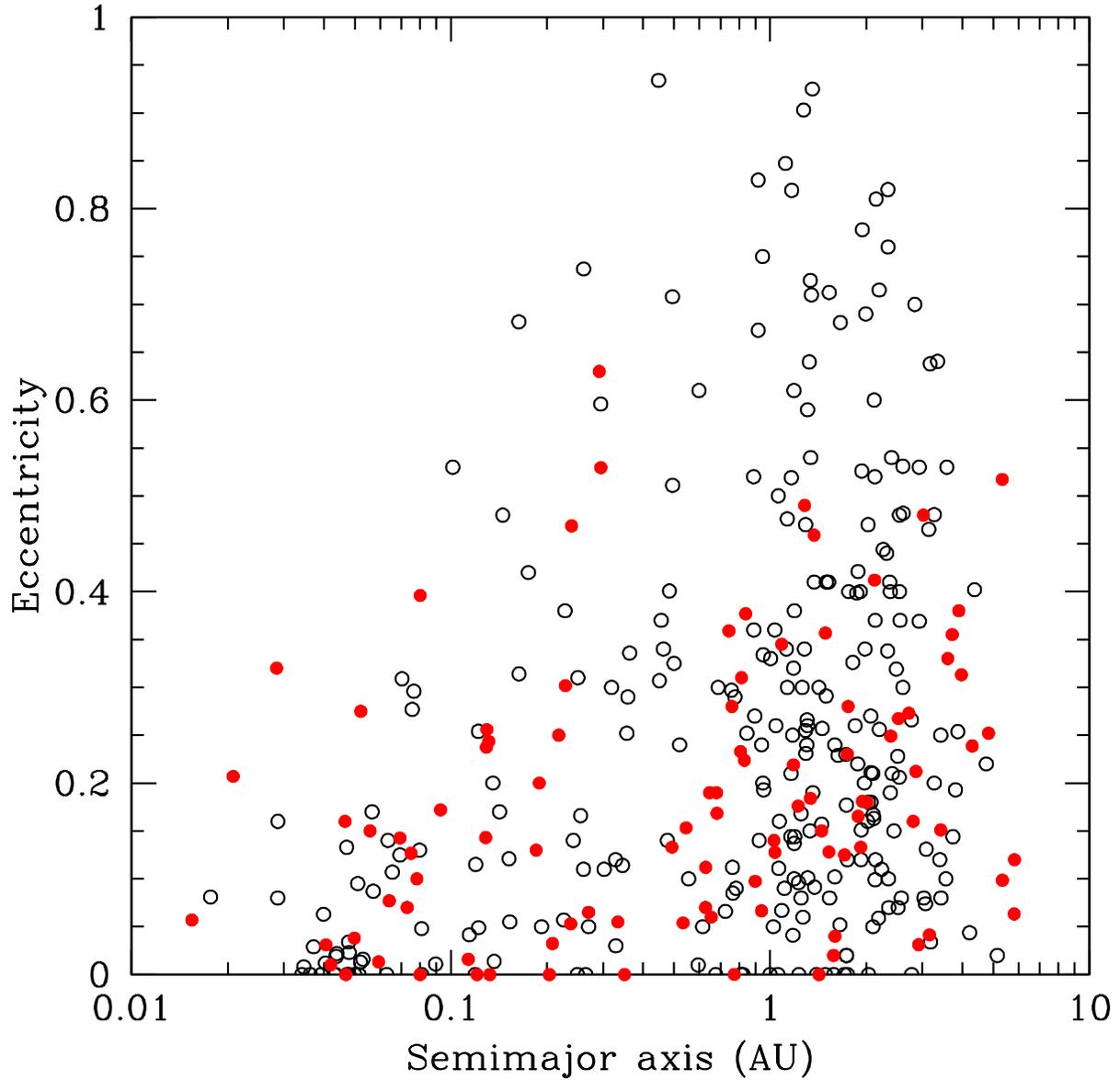}
\caption{Distribution of eccentricity versus semimajor axis for 
radial-velocity discovered planets in single systems (open circles) and 
multiple systems (filled circles). Planet data from the Exoplanet Orbit 
Database at exoplanets.org (2012 Jan 12).  Planets in multiple systems 
are marginally less eccentric than single planets. }
\label{compare}
\end{figure}

In both the HD~142 and HD~159868 systems, further monitoring has 
revealed additional planets, and the best-fit eccentricities of the 
previously known planets have significantly decreased.  
Figure~\ref{compare} shows the distribution of eccentricity versus 
semimajor axis for single planets (open circles) and multiple planets 
(filled circles).  A K--S test shows that there is only a 2.9\% 
probability that the eccentricities of single- and multiple-planet 
systems are drawn from the same distribution.  Previous analysis of the 
properties of multiple-planet systems have given the same result: that 
planets in multiple systems tend to have lower eccentricities than 
single planets \citep{wittenmyer09, wright09}.  To investigate whether 
this difference arises from an observational bias, we asked the 
following question: ``Is there a minimum threshold number of 
observations $N$ for which single and multiple planets have the same 
eccentricity distribution?'' We repeated the K--S tests on subsets of 
the known exoplanet data, including only those planets which have more 
than $N$ observations for a range of $N$, as shown in 
Table~\ref{eccdistrib}.  If there is a bias arising from the number of 
observations, then for larger $N_{obs}$, there would be no significant 
difference in the eccentricity distributions of single and multiple 
planet systems.  The K--S significance levels in Table~\ref{eccdistrib} 
do not show any consistent trend with $N_{obs}$.  We conclude that the 
difference in the eccentricity distributions of single and multiple 
planets is real and does not arise from the observational sampling.

Both HD\,142b and HD\,159868c have low-eccentricity orbits near one year 
about stars similar to the Sun.  This raises the question of the 
potential habitability (e.g. Horner \& Jones 2010) of terrestrial moons 
which may orbit these giant planets, or Trojan companions of those 
planets, a topic explored in detail by \citet{tinney11} for HD~38283b, a 
gas-giant planet in a one-year orbit.

While it is somewhat speculative to discuss the potential habitability 
of as-yet undiscovered moons or planet-mass Trojan companions of planets 
such as HD~142b and HD~159868c, it is important to note that, at least 
for our own Solar system, the capture of objects to such orbits is now 
considered a well established part of planetary formation and migration.  
In \citet{tinney11}, we provide an extensive review of the research into 
satellite formation and evolution, but it is worth reminding the reader 
of a few salient points. First, the combined mass of the largest 
satellites of the gas giant planets in our Solar system typically 
amounts to ~$2.5 \times 10^{-4}$ that of their host planet. In other 
words, it seems reasonable to expect that the most massive satellites of 
HD~142b could well be similar to, or somewhat more massive, than the 
Galilean satellites (since HD~142b is around 1.2 times the mass of 
Jupiter), whilst those of HD~159868c would likely be slightly less 
massive. As such, the regular satellites of those planets (assuming the 
same formation mechanism as the Galilean satellites) might well be 
somewhat too small to host sufficient atmosphere to allow liquid water 
on their surface. However, larger and more habitable satellites are 
clearly not beyond the bounds of possibility - particularly when one 
considers the possibility of the capture of massive irregular satellites 
(such as Neptune's moon Triton) during the course of the planet's 
migration \citep{JewShep05,JH07}.

A more promising alternative in the search for habitable exoplanets in 
these systems could be the capture of objects as Trojans of the planets 
in question.  Within the Solar system, it is known that objects can be 
temporarily captured as Trojans for long periods of time, even in the 
absence of planetary migration (e.g. Horner \& Evans 2006).  However, it 
is now widely accepted \citep{LH10, LH9, Morbi05} that the migration of 
the giant planets resulted in their capturing significant populations of 
Trojans.  In the case of Jupiter and Neptune, those Trojan populations 
were captured on orbits of sufficient stability that they have survived 
to the current day.  If either of HD\,142~b and HD\,159868~c was able to 
capture a sufficiently large planetary embryo as a Trojan during their 
inward migration to their current location, it is highly likely that 
such an object could remain trapped as a Trojan for the lifetime of the 
planetary system.  If such planet-mass Trojans exist in either system, 
they could well represent potentially habitable worlds.

\acknowledgements

JH and CGT gratefully acknowledge the financial support of the 
Australian government through ARC Grant DP0774000.  RW is supported by a 
UNSW Vice-Chancellor's Fellowship.  MT is supported by the RoPACS (Rocky 
Planets Around Cool Stars), a Marie Curie Initial Training Networks 
funded by the European Commission's Seventh Framework Programme.  JSJ 
acknowledges support through Fondecyt grant 3110004.  We gratefully 
acknowledge the UK and Australian government support of the 
Anglo-Australian Telescope through their PPARC, STFC and DIISR funding, 
and travel support from the Australian Astronomical Observatory and the 
Carnegie Institution of Washington.

We thank the ATAC for the generous allocation of telescope time which 
facilitated this detection.  This research has made use of NASA's 
Astrophysics Data System (ADS), and the SIMBAD database, operated at 
CDS, Strasbourg, France.


\begin{deluxetable}{lrr}
\tabletypesize{\scriptsize}
\tablecolumns{3}
\tablewidth{0pt}
\tablecaption{AAT Radial Velocities for HD 142}
\tablehead{
\colhead{JD-2400000} & \colhead{Velocity (\ms)} & \colhead{Uncertainty
(\ms)}}
\startdata
\label{142vels}
50830.95872  &     29.8  &    3.2  \\
51121.01944  &     36.9  &    3.6  \\
51385.31047  &     84.7  &    5.9  \\
51411.20252  &     75.5  &    6.3  \\
51473.08503  &     33.9  &    3.3  \\
51525.92509  &     54.0  &    4.0  \\
51526.95290  &     46.3  &    2.9  \\
51683.33138  &     86.4  &    3.8  \\
51743.27654  &     85.3  &    3.4  \\
51745.26417  &     90.7  &    5.5  \\
51767.26990  &     63.9  &    3.6  \\
51768.25417  &     66.1  &    3.2  \\
51828.06072  &     40.4  &    3.9  \\
51856.06429  &     12.0  &    5.5  \\
51856.92498  &     46.3  &    7.4  \\
51918.94072  &     55.5  &    3.8  \\
52061.29661  &     84.6  &    3.4  \\
52092.26831  &     70.0  &    3.3  \\
52093.28756  &     59.4  &    3.3  \\
52127.22295  &     34.6  &    4.1  \\
52128.15455  &     42.4  &    3.9  \\
52130.24335  &     35.4  &    3.6  \\
52151.21126  &     31.6  &    2.9  \\
52152.07857  &     36.5  &    3.7  \\
52154.15414  &     28.8  &    3.3  \\
52187.09998  &     34.9  &    2.9  \\
52188.03596  &     30.0  &    2.9  \\
52189.01990  &     22.3  &    3.0  \\
52190.00244  &     29.8  &    2.9  \\
52423.32977  &     58.3  &    3.1  \\
52425.33759  &     38.8  &    3.2  \\
52456.32088  &     15.8  &    3.5  \\
52477.24794  &     25.6  &    3.4  \\
52511.09845  &     37.8  &    3.4  \\
52654.91588  &     59.7  &    4.7  \\
52784.32722  &     36.4  &    3.4  \\
52857.23832  &     24.1  &    2.3  \\
52861.30634  &     24.1  &    5.1  \\
52946.03964  &     27.6  &    3.8  \\
53007.97796  &     35.1  &    3.9  \\
53041.95410  &     72.0  &    4.4  \\
53042.90882  &     64.9  &    4.9  \\
53215.27934  &     -7.3  &    2.6  \\
53243.28286  &      6.6  &    3.7  \\
53244.22676  &      0.4  &    3.8  \\
53246.09497  &      4.5  &    2.9  \\
53281.13767  &     -0.8  &    3.6  \\
53509.33429  &      4.5  &    2.0  \\
53516.33116  &      2.5  &    2.3  \\
53570.30140  &    -17.4  &    1.8  \\
53576.22037  &     -8.8  &    1.7  \\
53579.27530  &    -11.6  &    1.7  \\
53632.19723  &    -41.2  &    1.9  \\
53942.21660  &    -33.9  &    1.6  \\
54008.12359  &    -22.8  &    2.2  \\
54013.13830  &    -11.5  &    1.7  \\
54016.18494  &    -22.3  &    1.8  \\
54038.10169  &    -12.5  &    1.8  \\
54120.92520  &     22.7  &    2.1  \\
54255.26045  &    -32.8  &    2.5  \\
54334.13051  &    -43.4  &    2.5  \\
54374.14271  &    -24.4  &    2.1  \\
54428.94253  &     -3.0  &    1.7  \\
54780.10864  &     -3.4  &    1.5  \\
55076.28334  &    -58.7  &    2.1  \\
55101.11333  &    -19.8  &    4.2  \\
55170.91268  &     -7.3  &    2.0  \\
55376.32957  &    -89.9  &    3.8  \\
55377.32745  &    -86.0  &    2.7  \\
55401.20729  &    -59.7  &    2.3  \\
55428.24962  &    -16.4  &    4.0  \\
55457.13465  &    -16.3  &    2.5  \\
55518.99948  &     32.5  &    2.3  \\
55521.00990  &     10.7  &    2.6  \\
55751.31147  &    -31.3  &    3.6  \\
55757.24366  &    -34.3  &    2.7  \\
55786.25163  &      0.7  &    3.2  \\
55788.32904  &    -22.7  &    3.2  \\
55845.10610  &     14.7  &    2.6  \\
55874.03173  &     20.5  &    3.0  \\
55874.98064  &     15.4  &    3.6  \\
55897.91075  &      7.6  &    2.3  \\
\enddata
\end{deluxetable}

\begin{deluxetable}{lrr}
\tabletypesize{\scriptsize}
\tablecolumns{3}
\tablewidth{0pt}
\tablecaption{AAT Radial Velocities for HD 159868}
\tablehead{
\colhead{JD-2400000} & \colhead{Velocity (\ms)} & \colhead{Uncertainty
(\ms)}}
\startdata
\label{159vels}
52390.22780  &     29.1  &    1.4  \\
52422.14712  &     10.2  &    1.4  \\
52453.04281  &     11.9  &    1.5  \\
52456.07120  &     20.2  &    1.6  \\
52477.02060  &      8.1  &    1.5  \\
52711.26892  &    -29.3  &    2.9  \\
52747.25274  &    -34.7  &    1.5  \\
52751.26030  &    -34.4  &    1.4  \\
52786.09989  &    -59.4  &    1.3  \\
52858.95394  &    -25.3  &    1.3  \\
52942.93463  &    -17.4  &    1.9  \\
53214.09869  &      8.2  &    1.8  \\
53216.04487  &      9.1  &    1.4  \\
53242.96824  &     20.6  &    1.3  \\
53484.23874  &     31.4  &    1.3  \\
53486.16508  &     35.1  &    1.4  \\
53510.17399  &     18.3  &    1.4  \\
53521.19560  &     17.8  &    1.4  \\
53572.06973  &     34.9  &    1.3  \\
53631.89779  &     49.5  &    1.2  \\
53842.24103  &    -31.5  &    1.5  \\
53939.00914  &    -16.8  &    1.3  \\
53947.05674  &    -21.3  &    1.2  \\
54008.91766  &     -9.9  &    0.9  \\
54011.90938  &     -5.1  &    1.4  \\
54015.94916  &     -9.4  &    1.1  \\
54017.89861  &     -3.0  &    1.3  \\
54037.89603  &     -9.0  &    1.4  \\
54224.26216  &    -40.6  &    1.2  \\
54227.16135  &    -28.6  &    1.3  \\
54255.03411  &    -27.9  &    1.2  \\
54371.88740  &     25.7  &    1.1  \\
54553.20262  &     18.1  &    1.6  \\
54907.25523  &      4.8  &    1.7  \\
55101.91055  &      1.3  &    1.2  \\
55104.94555  &    -10.7  &    1.4  \\
55109.96044  &    -22.7  &    1.3  \\
55313.25667  &    -52.9  &    1.4  \\
55317.16029  &    -35.3  &    1.5  \\
55376.12410  &    -16.8  &    1.5  \\
55399.06643  &     -7.1  &    1.4  \\
55429.85887  &      6.8  &    1.4  \\
55456.87174  &      4.8  &    1.7  \\
55664.29971  &     19.6  &    1.3  \\
55692.24347  &     36.7  &    1.5  \\
55756.91450  &     74.4  &    1.9  \\
55786.07264  &     65.2  &    1.9  \\
\enddata
\end{deluxetable}

\begin{deluxetable}{lrr}
\tabletypesize{\scriptsize}
\tablecolumns{3}
\tablewidth{0pt}
\tablecaption{Keck Radial Velocities for HD 159868}
\tablehead{
\colhead{JD-2400000} & \colhead{Velocity (\ms)} & \colhead{Uncertainty
(\ms)}}
\startdata
\label{Keckvels}
54246.99415  &    -39.3  &    0.8  \\
54247.97421  &    -32.3  &    1.1  \\
54248.92666  &    -35.4  &    1.1  \\
54251.95192  &    -24.7  &    1.1  \\
54255.97421  &    -24.2  &    0.8  \\
54277.88461  &    -17.7  &    1.2  \\
54278.93840  &    -13.8  &    1.1  \\
54304.82235  &      2.3  &    0.8  \\
54305.83320  &      6.3  &    0.7  \\
54306.82686  &      0.0  &    1.0  \\
54307.86512  &      3.6  &    0.7  \\
54308.85011  &     -0.6  &    0.7  \\
54309.83390  &     -1.1  &    0.7  \\
54310.82821  &      6.4  &    0.7  \\
54311.82166  &      9.2  &    0.7  \\
54312.81832  &      8.5  &    0.7  \\
54313.79003  &      9.7  &    0.7  \\
54314.79189  &      8.9  &    0.7  \\
54335.72425  &     20.8  &    1.1  \\
54601.91517  &     24.8  &    1.1  \\
55024.86442  &     -2.8  &    0.6  \\
55049.83900  &    -11.9  &    0.8  \\
55052.81141  &     -0.7  &    0.6  \\
55260.15738  &    -60.9  &    1.0  \\
55369.98576  &     -8.6  &    0.6  \\
55409.84249  &    -10.9  &    0.6  \\
55462.73665  &     -2.1  &    0.7  \\
55638.13563  &     -1.6  &    0.8  \\
55665.10001  &     14.3  &    0.7  \\
55670.06654  &     14.8  &    0.7  \\
55720.04108  &     48.9  &    0.7  \\
55750.87325  &     56.1  &    1.3  \\
55825.74817  &     40.1  &    0.7  \\
55839.72241  &     35.0  &    1.2  \\
\enddata
\end{deluxetable}

\begin{deluxetable}{lll}
\tabletypesize{\scriptsize}
\tablecolumns{3}
\tablewidth{0pt}
\tablecaption{Stellar Parameters for HD 142 }
\tablehead{
\colhead{Parameter} & \colhead{Value} & \colhead{Reference}
 }
\startdata
\label{142star}
Spec.~Type & G1 IV & \citet{tinney02} \\
    & F7 V & \citet{gray06} \\
Mass (\Msun) & 1.15$\pm$0.10 & \citet{tinney02} \\ 
    & 1.232$^{+0.22}_{-0.16}$ & \citet{takeda07} \\
Distance (pc) & 25.7$\pm$0.3 & \citet{vl07} \\
$M_V$ & 3.66 &  \\
Radius (\Rsun) & 1.47$\pm$0.04 & \citet{vanbelle09} \\ 
    & 1.40$\pm$0.05 & \citet{takeda07} \\
    & 1.43$\pm$0.07 & \citet{lang80} \\
V sin $i$ (\kms) & 10.4$\pm$0.5 & \citet{vf05} \\
    & 10.35$\pm$0.50 & \citet{butler06} \\ 
log$R'_{HK}$ & -4.92 & \citet{tinney02} \\
    & -4.95 & \citet{jenkins06} \\
$[Fe/H]$ & 0.10$\pm$0.03 & \citet{vf05} \\
    & 0.09$\pm$0.05 & \citet{sousa08} \\
    & -0.02$\pm$0.07 & \citet{bond06} \\
    & 0.117$\pm$0.070 & \citet{gonzalez07} \\
    & -0.02$\pm$0.06 & \citet{ramirez07} \\
$T_{eff}$ (K) & 6403$\pm$65 & \citet{sousa08} \\
    & 6150$\pm$35 & \citet{bond06} \\
    & 6245$\pm$48 & \citet{ms11} \\
    & 6249 & \citet{vf05} \\
    & 6170 & \citet{randich99} \\
log $g$ & 4.62$\pm$0.07 & \citet{sousa08} \\
    & 4.26$^{+0.3}_{-0.2}$ & \citet{takeda07} \\
    & 4.19 & \citet{vf05} \\
    & 4.2 & \citet{randich99} \\
\enddata
\end{deluxetable}

\begin{deluxetable}{lll}
\tabletypesize{\scriptsize}
\tablecolumns{3}
\tablewidth{0pt}
\tablecaption{Stellar Parameters for HD 159868 }
\tablehead{
\colhead{Parameter} & \colhead{Value} & \colhead{Reference}
 }
\startdata
\label{159star}
Spec.~Type & G5 V & \citet{houk78} \\
Mass (\Msun) & 1.087$^{+0.032}_{-0.033}$ & \citet{takeda07} \\
   & 1.16$^{+0.27}_{-0.18}$ & \citet{vf05} \\ 
   & 0.919 & \citet{sousa08} \\
Distance (pc) & 52.7$\pm$3.0 & \citet{perryman97} \\
$M_V$ & 3.63 & \\
V sin $i$ (\kms) & 2.1 & \citet{vf05} \\ 
log$R'_{HK}$ & -4.96 & \citet{jenkins06} \\
$[Fe/H]$ & 0.00 & \citet{vf05} \\
   & -0.08$\pm$0.01 & \citet{sousa08} \\
$T_{eff}$ (K) & 5623 & \citet{vf05} \\
   & 5558$\pm$15 & \citet{sousa08} \\
log $g$ & 3.99$^{+0.05}_{-0.04}$ & \citet{takeda07} \\
   & 3.92 & \citet{vf05} \\
   & 3.96$\pm$0.02 & \citet{sousa08} \\
\enddata
\end{deluxetable}

\begin{deluxetable}{lr@{$\pm$}lr@{$\pm$}lr@{$\pm$}lr@{$\pm$}lr@{$\pm$}lr@{$\pm$}
lr@{$\pm$}l}
\tabletypesize{\scriptsize}
\tablecolumns{10}
\tablewidth{0pt}
\tablecaption{Keplerian Orbital Solutions }
\tablehead{
\colhead{Planet} & \multicolumn{2}{c}{Period} & \multicolumn{2}{c}{$T_0$}
&
\multicolumn{2}{c}{$e$} & \multicolumn{2}{c}{$\omega$} &
\multicolumn{2}{c}{K } & \multicolumn{2}{c}{M sin $i$ } &
\multicolumn{2}{c}{$a$ } \\
\colhead{} & \multicolumn{2}{c}{(days)} & \multicolumn{2}{c}{(JD-2400000)}
&
\multicolumn{2}{c}{} &
\multicolumn{2}{c}{(degrees)} & \multicolumn{2}{c}{(\ms)} &
\multicolumn{2}{c}{(\Mjup)} & \multicolumn{2}{c}{(AU)}
 }
\startdata
\label{planetparams}   
HD 142 b & 349.7 & 1.2 & 52683 & 26 & 0.17 & 0.06 & 327 & 26 & 
33.2 & 2.5 & 1.25 & 0.15 & 1.02 & 0.03 \\
HD 142 c & 6005 & 477 & 55954 & 223 & 0.21 & 0.07 & 250 & 20 &
55.2 & 3.0 & 5.3 & 0.7 & 6.8 & 0.5 \\
HD 159868 b & 1178.4 & 8.8 & 53435 & 56 & 0.01 & 0.03 & 350 & 171 &
38.3 & 1.1 & 2.10 & 0.11 & 2.25 & 0.03 \\
HD 159868 c & 352.3 & 1.3 & 53239 & 21 & 0.15 & 0.05 & 290 & 25 &
20.1 & 1.1 & 0.73 & 0.05 & 1.00 & 0.01 \\
\enddata
\end{deluxetable}


\begin{table}
\caption{The relative posterior probabilities of models 
$\mathcal{M}_{k}$ with $ k = 0, ..., 3$ Keplerian signals given the AAT 
data for HD\,142. The velocity jitter was also fitted as a free 
parameter in the model. }
\label{model_probabilities}
\begin{tabular}{lccc}
\hline \hline
$k$ & $P(\mathcal{M}_{k} | d)$ & Jitter (\ms) & RMS [ms$^{-1}$] \\
\hline
0 & $1.5 \times 10^{-48}$ & 39.7 & 39.1 \\
1 & $1.2 \times 10^{-32}$ & 23.7 & 22.5 \\
2 & $1.4 \times 10^{-8}$ & 11.2 & 11.2 \\
3 & $\sim$1 & 8.8 & 8.6 \\
\hline \hline
\end{tabular}
\end{table}


\begin{deluxetable}{lccc}
\tabletypesize{\scriptsize}
\tablecolumns{4}
\tablewidth{0pt}
\tablecaption{The three-planet solution of HD\,142 radial velocities. MAP 
estimates of the parameters and their 99\% Bayesian credibility 
sets.}
\tablehead{
\colhead{Parameter} & \colhead{Planet b} & \colhead{Planet c} & 
\colhead{Residual\tablenotemark{a}} \\
 }
\startdata
\label{HD142_parameters}
quit
$P$ [days] & 351.1 [348.3, 353.8] & 7900 [5500, 22200] & 108.39 [107.79, 109.00] \\
$e$ & 0.15 [0.00, 0.34] & 0.18 [0.00, 0.72] & 0.12 [0.00, 0.56] \\
$K$ [ms$^{-1}$] & 31.6 [26.1, 37.1] & 52.6 [40.0, 75.9] & 11.6 [6.1, 16.5] \\
$\omega$ [rad] & 4.3 [3.3, 6.2] & 3.2 [1.8, 4.9] & 4.7 [0, $2\pi$] \\
$M_{0}$ [rad]& 2.6 [1.4, 5.1] & 3.0 [0, $2\pi$] & 5.3 [0, $2\pi$] \\
$m_{p} \sin i$ [M$_{\rm Jup}$] & 1.21 [0.88, 1.54] & 5.5 [3.7, 11.6] & 0.30 [0.15, 0.45] \\
$a$ [AU] & 1.028 [0.915, 1.120] & 8.0 [6.0, 17.8] & 0.469 [0.418, 0.511] \\
\hline
$\gamma$ [ms$^{-1}$] & 8.6 [-5.6, 74.1] \\
$\sigma$ [ms$^{-1}$] & 8.8 [6.3, 11.8] \\
\enddata
\tablenotetext{a}{In the Bayesian analysis, we fit a third planet to 
illustrate the uncertainties in its parameters; we do not as yet claim a 
third planet in this system.}
\end{deluxetable}


\begin{table}
\caption{The relative posterior probabilities of models
$\mathcal{M}_{k}$ with $ k = 0, ..., 2$ Keplerian signals given the
AAT data for HD\,159868.}
\label{bayes159}
\begin{tabular}{lcc}
\hline \hline
$k$ & $P(\mathcal{M}_{k} | d)$ & RMS [ms$^{-1}$] \\
\hline
0 & $4.1 \times 10^{-23}$ & 28.9 \\
1 & $1.3 \times 10^{-13}$ & 16.0 \\
2 & $\sim 1.0$ & 6.6 \\
\hline \hline
\end{tabular}
\end{table}

\clearpage


\begin{deluxetable}{lllll}
\tabletypesize{\scriptsize}
\tablecolumns{5}
\tablewidth{0pt}
\tablecaption{Separations for HD 142 and its Stellar Companion}
\tablehead{
\colhead{JD-2400000} & \colhead{Separation (arcsec)} & \colhead{Angle 
(degrees)} & \colhead{Instrument} & \colhead{Reference}
}
\startdata
\label{nici}
53180.7 &   4.1$\pm$0.02   & 184.16$\pm$0.18 & VLT-NACO & \citet{egg07} \\
53180.8 &  4.11$\pm$0.02   & 184.16$\pm$0.18 & VLT-NACO & \citet{egg07} \\
53316   &   4.1$\pm$0.02   & 184.18$\pm$0.18 & VLT-NACO & \citet{egg07} \\
53712.7 &   4.08$\pm$0.02   & 184.13$\pm$0.29 & VLT-NACO & \citet{egg07} \\
55391.43 &  3.965$\pm$0.013  & 184.47$\pm$0.26 & Gemini-NICI & This work \\
\enddata
\end{deluxetable}

\begin{deluxetable}{llll}
\tabletypesize{\scriptsize}
\tablecolumns{4}
\tablewidth{0pt}
\tablecaption{Eccentricity Distributions for Planets in Single and 
Multiple Systems}
\tablehead{
\colhead{Filter} & \colhead{K--S Probability\tablenotemark{a}} & 
\colhead{$N_{single}$} & \colhead{$N_{multiple}$}
 }
\startdata
\label{eccdistrib}
All $N_{obs}$ & 0.029 & 268 & 98 \\
$N_{obs}>40$ & 0.006 & 138 & 90 \\
$N_{obs}>50$ & 0.040 & 99 & 88 \\
$N_{obs}>60$ & 0.005 & 64 & 81 \\
$N_{obs}>70$ & 0.088 & 53 & 76 \\
$N_{obs}>80$ & 0.519 & 45 & 66 \\
$N_{obs}>90$ & 0.186 & 37 & 62 \\
$N_{obs}>100$ & 0.196 & 30 & 60 \\
$N_{obs}>110$ & 0.082 & 24 & 49 \\
$N_{obs}>120$ & 0.039 & 20 & 43 \\
$N_{obs}>130$ & 0.063 & 18 & 43 \\
$N_{obs}>140$ & 0.504 & 12 & 40 \\
\enddata
\tablenotetext{a}{Probability that the two samples are drawn from the same
distribution.}
\end{deluxetable}


\begin{thebibliography}{}

\bibitem[Anglada-Escud{\'e} et al.(2012)]{ang12} 
Anglada-Escud{\'e}, G., Arriagada, P., Vogt, S.~S., et al.\ 2012, \apjl, 
751, L16

\bibitem[Anglada-Escud{\'e} et al.(2010)]{ang10} Anglada-Escud{\'e}, G., 
L{\'o}pez-Morales, M., \& Chambers, J.~E.\ 2010, \apj, 709, 168

\bibitem[Bailey et al.(2009)]{bailey09} Bailey, J., Butler, R.~P., 
Tinney, C.~G., Jones, H.~R.~A., O'Toole, S., Carter, B.~D., \& Marcy, 
G.~W.\ 2009, \apj, 690, 743

\bibitem[Bean et al.(2008)]{bean08} Bean, J.~L., McArthur, B.~E., 
Benedict, G.~F., \& Armstrong, A.\ 2008, \apj, 672, 1202

\bibitem[\protect\astroncite{Bond et al.}{2006}]{bond06} Bond, J. C., 
Tinney, C. G., Butler, R. P., et al. 2006, MNRAS, 370, 163


\bibitem[\protect\citeauthoryear{Butler et al.}{1996}]{BuMaWi96}
    Butler, R.~P., Marcy, G.~W., Williams, E., McCarthy, C.,
    Dosanjh, P., \& Vogt, S.~S. 1996, \pasp, {108}, 500

\bibitem[\protect\astroncite{Butler et al.}{2006}]{butler06} Butler, R. 
P., Wright, J. T., Marcy, G. W., et al. 2006, ApJ, 646, 505

\bibitem[Chambers (1999)]{chambers99} Chambers, J. E.\ 1999, MNRAS, 304, 
793

\bibitem[Cochran et al.(2007)]{cochran07} Cochran, W.~D., Endl, M., 
Wittenmyer, R.~A., \& Bean, J.~L.\ 2007, ApJ, 665, 1407

\bibitem[\protect\citeauthoryear{Diego et~al.}{1991}]{diego:90}
    Diego, F., Charalambous, A., Fish, A.~C., \& Walker, D.~D. 1990,  
Proc. Soc. Photo-Opt. Instr. Eng., 1235, 562

\bibitem[Eggenberger et al.(2007)]{egg07} Eggenberger, A., 
Udry, S., Chauvin, G., et al.\ 2007, \aap, 474, 273

\bibitem[Fabrycky et al.(2012)]{fab12} Fabrycky, D.~C., Ford, E.~B., 
Steffen, J.~H., et al.\ 2012, \apj, 750, 114

\bibitem[Ford \& Gregory(2007)]{ford2007} Ford, E.~B., \& Gregory, 
P.~C.\ 2007, Statistical Challenges in Modern Astronomy IV, 371, 189

\bibitem[\protect\astroncite{Gonzalez \& Laws}{2007}]{gonzalez07} 
Gonzalez, G. \& Laws, C. 2007. MNRAS, 378, 1141

\bibitem[\protect\astroncite{Gray}{2006}]{gray06} Gray, R. O., Corbally, 
C. J., Garrison, R. F., et al. 2006, ApJ, 132, 161

\bibitem[Gregory \& Fischer(2010)]{gf10} Gregory, P.~C., 
\& Fischer, D.~A.\ 2010, \mnras, 403, 731

\bibitem[\protect\astroncite{Haario et al.}{2001}]{haario2001} Haario, 
H., Saksman, E., \& Tamminen, J. 2001, Bernoulli, 7, 223




\bibitem[Horner \& Evans (2006)]{Horner06} Horner, J., \& Evans, N. W., 
2006, MNRAS, 367, L20

\bibitem[Horner \& Jones(2010)]{HornerJones10} Horner, J., \& 
Jones, B.~W.\ 2010, International Journal of Astrobiology, 9, 273

\bibitem[Horner et al.(2011)]{horner11} Horner, J., Marshall, J.~P., 
Wittenmyer, R.~A., \& Tinney, C.~G.\ 2011, MNRAS, 416, L11


\bibitem[Horner et al.(2012)]{HUAqr2} Horner, J., Wittenmyer, R.~A., 
Marshall, J.~P., Tinney, C.~G., \& Butters, O.~W.\ 2012, Proceedings of 
the 11th Australian Space Science Conference, in press.

\bibitem[Houk(1978)]{houk78} Houk, N.\ 1978, Ann Arbor : Dept.~of 
Astronomy, University of Michigan : distributed by University Microfilms 
International, 1978-,

\bibitem[Isaacson \& Fischer(2010)]{if10} Isaacson, H., 
\& Fischer, D.\ 2010, \apj, 725, 875

\bibitem[Jefferys et al.(1987)]{jefferys87} Jefferys, W.~H., 
Fitzpatrick, M.~J., \& McArthur, B.~E.\ 1987, Celestial Mechanics, 41, 
39

\bibitem[Jenkins et al.(2006)]{jenkins06} Jenkins, J.~S., Jones, 
H.~R.~A., Tinney, C.~G., et al.\ 2006, \mnras, 372, 163

\bibitem[Jewitt \& Sheppard(2005)]{JewShep05} Jewitt, D., \& Shepard, S., 
2005, Space Science Reviews, 116, 441

\bibitem[Jewitt \& Haghighipour(2007)]{JH07} Jewitt, D., \& Haghighipour, 
N., 2007, Annual Review of Astronomy and Astrophysics, 45, 261

\bibitem[Johnson et al.(2010)]{johnson10a} Johnson, J.~A., Aller, K.~M., 
Howard, A.~W., \& Crepp, J.~R.\ 2010, \pasp, 122, 905

\bibitem[Jones et al.(2010)]{jones10} Jones, H.~R.~A., et al.\ 2010, 
\mnras, 403, 1703

\bibitem[Kjeldsen \& Bedding(1995)]{kb95} Kjeldsen, H., 
\& Bedding, T.~R.\ 1995, \aap, 293, 87

\bibitem[K{\" u}rster et al.(1997)]{kurster97} K{\" u}rster, M., 
Schmitt, J.~H.~M.~M., Cutispoto, G., \& Dennerl, K.\ 1997, \aap, 320, 
831

\bibitem[\protect\astroncite{Lang}{1980}]{lang80} Lang, K. R. 1980, 
Astrophysical formulae. A compendium for the physicist and 
astrophysicist, XXIX, 783pp. Springer-Verlag Berlin Heidelberd New York

\bibitem[Lykawka et al.(2009)]{LH9} Lykawka, P.~S., Horner, J., Jones, 
B.~W., \& Mukai, T.\ 2009, \mnras, 398, 1715

\bibitem[Lykawka \& Horner(2010)]{LH10} Lykawka, P.~S., 
\& Horner, J.\ 2010, \mnras, 405, 1375

\bibitem[\protect\astroncite{Malyuto \& Shvelidze}{2011}]{ms11} Malyuto, 
V. \& Shvelidze, T. 2011, Baltic Astronomy, 20, 91

\bibitem[Marshall et al.(2010)]{marshall10} Marshall, J., Horner, J., \& 
Carter, A.\ 2010, International Journal of Astrobiology, 9, 259

\bibitem[Mayor et al.(2011)]{mayor11} Mayor, M., Marmier, M., Lovis, C., 
et al.\ 2011, arXiv:1109.2497

\bibitem[McCarthy et al.(2004)]{mccarthy04} McCarthy, C., Butler, R.~P., 
Tinney, C.~G., Jones, H.~R.~A., Marcy, G.~W., Carter, B., Penny, A.~J., 
\& Fischer, D.~A.\ 2004, \apj, 617, 575

\bibitem[Meschiari et al.(2011)]{mes11} Meschiari, S., Laughlin, G., 
Vogt, S.~S., et al.\ 2011, \apj, 727, 117

\bibitem[Morbidelli et al.(2005)]{Morbi05} Morbidelli, A., Levison, 
H.~F., Tsiganis, K., \& Gomes, R.\ 2005, \nat, 435, 462

\bibitem[O'Toole et al.(2007)]{otoole07} O'Toole, S.~J., Butler, R.~P., 
Tinney, C.~G., et al.\ 2007, \apj, 660, 1636

\bibitem[O'Toole et al.(2009)]{otoole09} O'Toole, S.~J., Jones, 
H.~R.~A., Tinney, C.~G., et al.\ 2009, \apj, 701, 1732

\bibitem[Pepe et al.(2011)]{pepe11} Pepe, F., Lovis, C., 
S{\'e}gransan, D., et al.\ 2011, \aap, 534, A58

\bibitem[Perryman et al.(1997)]{perryman97} Perryman, M.~A.~C., et al.\
1997, \aap, 323, L49

\bibitem[Poveda et al.(1994)]{poveda94} Poveda, A., Herrera, M.~A., 
Allen, C., Cordero, G., \& Lavalley, C.\ 1994, RMxAA, 28, 43

\bibitem[Raghavan et al.(2006)]{rag06} Raghavan, D., Henry, T.~J., 
Mason, B.~D., et al.\ 2006, \apj, 646, 523

\bibitem[\protect\astroncite{Ram\'irez et al.}{2007}]{ramirez07} 
Ram\'irez, I., Allende Prieto, C., \& Lambert, D. L. 2007, A\&A, 465, 
271

\bibitem[Randich et al.(1999)]{randich99} Randich, S., Gratton, R., 
Pallavicini, R., Pasquini, L., \& Carretta, E.\ 1999, \aap, 348, 487

\bibitem[Rivera et al.(2005)]{rivera05} Rivera, E.~J., Lissauer, J.~J., 
Butler, R.~P., et al.\ 2005, \apj, 634, 625

\bibitem[Robertson et al.(2012)]{Texan} Robertson, P., Endl, M., 
Cochran, W.~D., et al.\ 2012, \apj, 749, 39

\bibitem[Sousa et al.(2008)]{sousa08} Sousa, S.~G., Santos, 
N.~C., Mayor, M., et al.\ 2008, \aap, 487, 373

\bibitem[Takeda et al.(2007)]{takeda07} Takeda, G., Ford, E.~B., Sills, 
A., Rasio, F.~A., Fischer, D.~A., \& Valenti, J.~A.\ 2007, \apjs, 168, 
297

\bibitem[Tinney et al.(2002)]{tinney02} Tinney, C.~G., Butler, R.~P., 
Marcy, G.~W., et al.\ 2002, \apj, 571, 528

\bibitem[Tinney et al.(2011)]{tinney11} Tinney, C.~G., Wittenmyer, 
R.~A., Butler, R.~P., Jones, H.~R.~A., O'Toole, S.~J., Bailey, J.~A., 
Carter, B.~D., \& Horner, J.\ 2011, ApJ, 732, 31

\bibitem[Tuomi(2011)]{tuomi11} Tuomi, M.\ 2011, \aap, 528, L5

\bibitem[Tuomi et al.(2012)]{tuomi12} Tuomi, M., et al.\ 2012, \mnras, 
submitted

\bibitem[\protect\astroncite{Tuomi \& Kotiranta}{2009}]{tuomi2009} 
Tuomi, M. \& Kotiranta, S. 2009, A\&A, 496, L13

\bibitem[\protect\astroncite{Tuomi}{2011}]{tuomi2011} Tuomi, M. 2011, 
A\&A, 528, L5

\bibitem[\protect\astroncite{Tuomi et al.}{2011}]{tuomi2011b} Tuomi, M., 
Pinfield, D., \& Jones, H. R. A. 2011, A\&A, 532, A116

\bibitem[Udry et al.(2003)]{udry03} Udry, S., Mayor, M., \& Queloz, D.\ 
2003, Scientific Frontiers in Research on Extrasolar Planets, 294, 17

\bibitem[Udry et al.(2006)]{udry06} Udry, S., Mayor, M., Benz, W., et 
al.\ 2006, \aap, 447, 361

\bibitem[Valenti \& Fischer(2005)]{vf05} Valenti, J.~A., \& Fischer, 
D.~A.\ 2005, \apjs, 159, 141

\bibitem[\protect\astroncite{van Belle \& von Braun}{2009}]{vanbelle09} 
van Belle, G. T. \& von Braun, K. 2009, ApJ, 694, 1085

\bibitem[van Leeuwen(2007)]{vl07} van Leeuwen, F.\ 2007, \aap, 474, 653

\bibitem[\protect\citeauthoryear{Valenti et~al.} {1995}]{val:95}
    Valenti, J.~A., Butler, R.~P. \& Marcy, G.~W. 1995,
    \newblock { PASP, } {107}, 966.

\bibitem[Vogt et al.(1994)]{vogt94} Vogt, S.~S., Allen, S.~L., Bigelow, 
B.~C., et al.\ 1994, \procspie, 2198, 362

\bibitem[Vogt et al.(2010)]{61vir} Vogt, S.~S., et al.\ 2010, \apj, 708, 
1366

\bibitem[Wittenmyer et al.(2009)]{wittenmyer09} Wittenmyer, R.~A., Endl, 
M., Cochran, W.~D., Levison, H.~F., \& Henry, G.~W.\ 2009, \apjs, 182, 
97

\bibitem[Wittenmyer et al.(2011a)]{jupiters} Wittenmyer, R.~A., Tinney, 
C.~G., O'Toole, S.~J., Jones, H.~R.~A., Butler, R.~P., Carter, B.~D., \& 
Bailey, J.\ 2011a, \apj, 727, 102

\bibitem[Wittenmyer et al.(2011b)]{47205paper} Wittenmyer, R.~A., Endl, 
M., Wang, L., et al.\ 2011b, \apj, 743, 184

\bibitem[Wittenmyer et al.(2012)]{HUAqr} Wittenmyer, 
R.~A., Horner, J., Marshall, J.~P., Butters, O.~W., \& Tinney, C.~G.\ 
2012, \mnras, 419, 3258

\bibitem[Wright(2005)]{wright05} Wright, J.~T.\ 2005, \pasp, 117, 657

\bibitem[Wright et al.(2009)]{wright09} Wright, J.~T., Upadhyay, S., 
Marcy, G.~W., et al.\ 2009, \apj, 693, 1084

\end{thebibliography}
\end{document}